\documentclass[a4paper,fleqn,usenatbib]{mnras}

\usepackage{newtxtext,newtxmath}     
\usepackage{changes}
\usepackage{comment}
\usepackage{color}

\usepackage[T1]{fontenc}
\usepackage{ae,aecompl}

\usepackage{graphicx}	
\usepackage{amsmath}	
\usepackage{amssymb}	

\newcommand{\ip}{{$i$-process }}
\newcommand{\spr}{{$s$-process}}
\newcommand{\isotope}[2]{${}^{#1}$#2}
\newcommand{\msun}{\mbox{$\mathrm{M_{\odot}}$}}
\newcommand{\rsun}{\mbox{$\mathrm{R_{\odot}}$}}
\newcommand{\lsun}{\mbox{$\mathrm{L_{\odot}}$}}
\newcommand{\mdot}{\mbox{$\dot{\mathrm{M}}$}}
\newcommand{\mdota}{\mbox{$\dot{\mathrm{M}}_{\mathrm{accr}}$}} 
\newcommand{\mdotd}{\mbox{$\dot{\mathrm{M}}_{\mathrm{tran}}$}} 
\newcommand{\pyr}{\mbox {{\rm yr$^{-1}$}}}
\newcommand{\myr}{\mbox {~${\rm M_{\odot}~yr^{-1}}$}}
\newcommand{\mwd}{\mbox {{\rm M$_{\rm WD}$}}}

\newcommand{\gcc}{\mbox {{\rm g~cm$^{-3}$}}}
\newcommand{\msunyrm}{\mbox{$\mathrm{M_\odot~yr^{-1}}$}}
\newcommand{\nean}{\mbox{$\mathrm{{^{22}Ne}(\alpha,n){^{25}Mg}}\  $}}
\newcommand{\am}{\mbox {AM~CVn }}
\def\apgt{\ {\raise-.5ex\hbox{$\buildrel>\over\sim$}}\ } 
\def\aplt{\ {\raise-.5ex\hbox{$\buildrel<\over\sim$}}\ } 
\def\sval{Sov. Astron. Lett.} 

\title[Nucleosynthesis in the extreme S102+030 binary system]{He-accreting WD: Nucleosynthesis in the extreme binary system (1.02+0.30)~\msun}

\author[L. Piersanti et al.]{
Luciano Piersanti,$^{1,2}$\thanks{E-mail: luciano.piersanti@inaf.it}
Lev R.Yungelson,$^{3}$
Sergio Cristallo$^{1,2}$
and Amedeo Tornamb\'e$^{4}$
\\
$^{1}$INAF - Osservatorio Astronomico d'Abruzzo, via Mentore Maggini, snc, 64100 Teramo, IT\\
$^{2}$INFN - Sezione di Perugia, Via A. Pascoli snc, 06123 Perugia, IT\\
$^{3}$Institute of Astronomy of the Russian Academy of Sciences, 48 Pyatnitskaya Str., 119017 Moscow, Russia\\
$^{4}$INAF-Osservatorio Astronomico di Roma, via Frascati, 33, 00040 Monte Porzio Catone, IT
}

\date{Accepted XXX. Received YYY; in original form ZZZ}

\pubyear{2018}

\begin{document}
\label{firstpage}
\pagerange{\pageref{firstpage}--\pageref{lastpage}}
\maketitle

\begin{abstract}
We investigate the evolutionary properties of AM~CVn stars with massive white dwarf donors and 
accretors. As a representative of them we consider a binary initially composed 
by a 0.30\msun\, He WD  and a 1.02\msun\, CO WD.
We evaluate the time-dependent mass transfer rate from the donor and compute the evolution of the 
accretor, accounting for the effects of mass exchange on the evolution of orbital 
parameters. We model the thermal response of the accreting CO WD with the FUN evolutionary code 
coupled to a full nuclear network, from H to Bi, including more than 700 isotopes linked 
by about 1000 nuclear processes.
We find that accretors in these systems evolve through the stages of  steady He-burning and mild and strong He-flashes 
and become at the end CO WDs capped by a massive ($\sim0.1$\msun) He-rich buffer. During He-flashes (both mild and strong) the temperature in the He-shell 
increases above $3\times 10^8$ K, so that the \nean reaction becomes efficient and $n$-rich isotopes 
can be produced. During the RLOF episodes triggered by strong non-dynamical He-flashes matter 
enriched in $\alpha$-elements and $n$-rich isotopes is ejected, polluting the interstellar medium.
Our results strongly suggest that massive AM CVn systems with WD donors do not experience a 
final very strong dynamical He-flash driving an explosive event like SN .Ia. Though the ejected 
matter is highly enriched in heavy isotopes, the relative contribution of massive AM CVn systems 
to the Galactic chemical evolution is, most probably, negligible due to their expected paucity.
\end{abstract}




\section{Introduction}
\label{sec:intro}

This paper continues a series of studies of helium accretion onto carbon-oxygen 
white dwarfs 
\citep[][henceforth, Papers~I and II, respectively]{2014MNRAS.445.3239P,piersanti2015}.
In particular, in Paper~II we considered time-dependent accretion in modeling 
ultracompact cataclysmic variables of \am type  with helium white dwarf (WD) donors and CO WD accretors, also dubbed ``interacting double-degenerates'' (IDD), 
in which the rate of mass-transfer is set by angular momentum loss via emission of gravitational waves (GW).
Fifty-six \am stars were known at the time of writing 
\citep{2018arXiv181006548R}. But we recall that it is suggested that, in addition to systems with WD donors, other two varieties
of \am stars may exist, harbouring either a low-mass helium star donor \citep{skh86} or a remnant 
of a strongly evolved main-sequence star \citep{tfey85}.

A detailed description of observational properties,  formation and evolution
of \am stars  can be found, e.g., in
\citet{nelemans_amcvn05,2009CQGra..26i4030N,2010PASP..122.1133S,2018arXiv181006548R}. 
We also refer the reader to \citet{mns04} for discussion of stability of mass-exchange in these binaries, Paper~I for investigation of regimes of He-burning at the surface of accreting CO WD in general  and to Paper~II for a study of some details of unstable He-burning in \am stars with variable mass-transfer rate.

Like in Paper II, the mass-transfer rate is determined by adopting the mass-radius (M-R) relation for cold degenerate stars.
We use an approximation to \citet{zs69} M-R relation devised by Eggleton and published in \citet{vr88}. The usage of this approximation 
may be justified by the following circumstances. Typical delay between formation of
detached  He+CO WD pairs and semidetached \am stars is $\sim$~Gyr \citep{ty96}. On the other hand, e.g., \citet{2007MNRAS.382..779P} have shown that the radii of low-mass helium WD remnants of mass-losing stars reach asymptotical values close to those given by 
M-R relation of \citeauthor{zs69} in the similar timescale. As well,
mass-transfer rates based on the latter M-R relation, excellently agree with those, based on M-R relation for finite, but low, entropy WDs suggested by 
\citet{2006ApJ...640..466B}.\footnote{Note, in the analysis of stability of mass-transfer in \am stars by \citet{mns04}, M-R relation published by  \citet{vr88} was used.}            

We remind also
 that, as found first by \citet{nom82a} and, most lately, e.g.,  in our Paper I
and  by \citet{2015ApJ...807...74B,2017A&A...604A..31W},
He-burning may occur in the following regimes (in the order of increasing accretion rate \mdota).
\begin{itemize}
\item Detonation: thermonuclear burning is initiated in degenerate layer close to CO/He interface and, 
if heating timescale by nuclear reactions is shorter than dynamical timescale of the He-burning layer, 
an explosion may be triggered. 
\item Strong flashes (SF): nuclear burning starts in degenerate matter, 
heating timescale does not become shorter than dynamical timescale, but released
energy is sufficient for ejection of a part or even entire accumulated He-layer, depending on the 
mass of accretor and \mdota. 
\item Mild flashes (MF): thermonuclear burning occurs in non-degenerate matter, energy release is not
sufficient for ejection of matter.
\item Steady-state (SS) burning:  WD burns accreted He into CO-mixture at a rate which is 
slightly lower than \mdota.
\item RG-regime: accretion rate \mdota\ is higher than accretor may burn in steady-state regime and 
excess of the matter forms 
an extended (red-giant-like) envelope around the WD or the envelope overflows Roche lobe of the WD.
\end{itemize} 
In Paper II we studied evolution of binaries with initial accretor+donor mass combinations 
(0.6+0.17)\msun,
(0.92+0.15)\msun, (1.02+0.2)\msun, which were suggested by population synthesis model of
\citet{npv+01}. 
In \am binaries with WD donors, \mdota\ decreases with time and, for the systems studied in Paper II,
He-burning proceeded through SS-, MF-, SF-regimes, but never became dynamic, as hypothesized
by \citet{bildsten2007} for the ``last'' flash, occurring at the lowest \mdota.
In both Paper II and the present work we neglected the effects of dynamical friction in the 
expanded envelope after the  Roche lobe overflow (RLOF) triggered by He-flashes 
and we assumed that the mass transfer remains dynamically stable. At variance, 
\citet{shen2015} suggested that such an occurrence leads inevitably to the merging of the two 
degenerate components.

The most massive system studied in Paper~II had initial masses of components
1.02 and 0.2\msun, well inside the region where,
under assumption of short time-scale of tidal synchronization in the binary, 
stable mass-transfer at a rate below the Eddington one is possible 
\citep{mns04}.
The system  we study  here -- (1.02+0.3)\msun, {\it aka} S102+030 --
is just at the border of stable/unstable mass-transfer. 
Binary population synthesis model for \am\ stars \citep{npv+01} shows that the 
systems with such combinations of components masses may form, though,
we clearly recognize that the results of population synthesis for any class of objects 
depends on specific set of accepted parameters, especially, as in the case of \am\ stars, 
on the assumptions on common envelope evolution.

Like in less massive  S102+020 system, WD in S102+030 experiences He-burning in 
SS, MF, and SF regimes listed above. Due to the larger mass of the accretor, 
during the evolution in the SF burning regime, the S102+030 system 
experiences He-flashes so strong that 
the temperature in the He-burning shell exceeds $3\times 10^8$ K. In this case, 
$\alpha$-captures on \isotope{22}{Ne} become efficient, leading to the release 
of neutrons via \nean reactions. We recall that the \isotope{22}{Ne} abundance in the He-rich layer 
is mainly determined by successive $\alpha$-captures by \isotope{14}{N} present in the accreted 
matter. The latter is resulting from the conversion of the initial CNO group elements into \isotope{14}{N} 
via H-burning in the He WD progenitor. This implies that in the He WD which descends from the 
primary component of a  
binary system with solar chemical composition (see Papers I and II), abundance
of \isotope{22}{Ne} is as large as $2.08\times 10^{-2}$ by mass fraction.\footnote{Hereafter 
we address the ``mass fraction abundance'' of a given isotope as ``abundance''.} 
Neutrons delivered by the \nean reaction are captured by heavy elements, releasing, on average, 
Q$\sim$5 MeV per each reaction, which is comparable to the energy contribution coming from 
$3\alpha$-reaction (Q=7.275 MeV) and $\alpha$-capture on the 
\isotope{12}{C} (Q=7.162 MeV).
Moreover, $n$-captures could trigger a peculiar nucleosynthesis both via \ip and 
weak \spr\ (see  \S\ref{sec:nuc}). 

The paper is organized as follows. In \S\ref{sec:nuc} we describe the possible nucleosynthesis path in 
hydrogen-deficient matter under physical conditions encountered in massive \am stars and 
we individuate the correct nuclear network to account for both the energy contribution and 
the chemical evolution triggered by the \nean reaction. 
In \S\ref{sec:beg} we briefly consider the initiation of mass-transfer in (1.02+0.3)\msun\ system, which differs from
the one in the systems with less massive donors.
In \S\ref{sec:ph_evol} an overview of the evolution of the (1.02+0.3)\msun\ binary in SF-regime is presented.
In \S\ref{sec:che_evol} an example of nucleosynthesis during particular strong thermal pulse is considered in detail
and overall nucleosynthesis path of the system is presented.
Discussion and conclusions follow in \S\ref{sec:discus}.

\begin{table*}  
	\centering
\caption{Selected physical properties of the system S102+030 during the first He-flash episode. The various epochs are:
         A - onset of the flash-driven convective shell; 
         B - ignition of He-burning;
         C - maximum luminosity of the He-burning shell;
         D - beginning of the RLOF episode; 
         E - disappearance of flash-driven convective shell;
         F - resumption of the mass transfer from the donor;
         G - resumption of effective mass deposition onto accretor.
         For each epoch we list 
         the time elapsed between two successive epochs  $\mathrm{\Delta t}$ in yr,  
         the total mass of the accretor $\mathrm{M_{acc}}$ and of the donor $\mathrm{M_{don}}$ in \msun, 
         the separation of components $a$ in $10^{-2}$\rsun, 
         the effective temperature $\mathrm{T_{eff}}$ and luminosity $\mathrm{L/L_\odot}$,
         mass coordinate of the He-burning shell $\mathrm{M_{He}}$ in \msun, the density 
         $\mathrm{\rho_{He}}$\ in $\mathrm{10^3 g\cdot cm^{-3}}$,
          the temperature $\mathrm{T_{He}}$\ in $10^8$ K,  and luminosity $\mathrm{L_{He}/\lsun}$ of the He-burning shell.
         The $\mathrm{\Delta t}$ for epoch A refers to the time from RLOF by donor WD to the onset of the 
          flash-driven convective shell in accreting WD. 
           } 
\label{tab1} 
\centering 
  \begin{tabular}{l r r r r r r r }
   \hline\hline
{\it } & A & B & C & D & E & F & G \\
   \hline
$\mathrm{\Delta t}$ (yr)          &  26.632 &    0.071 &    0.053 &    1.177 &    6.085 &    0.686 &    7.120 \\
$\mathrm{M_{acc}/M_\odot}$        &1.021028 & 1.021030 & 1.021031 & 1.021056 & 1.020784 & 1.020760 & 1.020678 \\
$\mathrm{M_{don}/M_\odot}$        &0.299438 & 0.299437 & 0.299436 & 0.299411 & 0.299406 & 0.299406 & 0.299329 \\
$a\mathrm{\ (in\ 10^{-2}R_\odot)}$& 6.1519  &  6.1519  &  6.1519  &  6.1523  &  6.1520  &  6.1519  &  6.1528  \\
$\mathrm{\log(T_{eff})}$          &  5.746  &   5.745  &   5.698  &   5.675  &   5.681  &   5.684  & 5.703\\
$\mathrm{\log(L/L_\odot)}$        &  3.916  &   3.914  &   3.835  &   4.608  &   4.632  &   4.645 & 4.713 \\
$\mathrm{M_{He}/M_\odot}$         & 1.02061 &  1.02057 &  1.02051 &  1.02029 &  1.02017 &  1.02007 & 1.02048 \\
$\mathrm{\rho_{He}}$              &  8.358  &   7.700  &   3.888  &   5.210  &   4.896  &   7.226  &   1.154  \\
$\mathrm{T_{He}}$                 &  1.979  &   2.288  &   3.350  &   2.934  &   3.007  &   2.890  &   2.340  \\
$\mathrm{\log(L_{He}/L_\odot)}$   &  5.026  &   5.918  &   7.154  &   5.650  &   4.501  &   4.440  &   4.196  \\
  \hline
  \end{tabular}
\end{table*}
\section{Nucleosynthesis via {$\lowercase{n}$}-capture processes}
\label{sec:nuc}	

Elements heavier than iron are mainly synthesized by the slow neutron capture process 
(the \spr) and the rapid neutron capture process (the $r$-process). Those processes, 
theorized 60 years ago by \citet{b2fh1957}, are characterized by either low 
($n_n\sim 10^6\div 10^7$ cm$^{-3}$ for the \spr) or very large neutron
densities ($n_n> 10^{20}$ cm$^{-3}$ for the $r$-process). 

The $r$-process produces neutron-rich isotopes very far from the $\beta$-stability valley. 
It was originally thought to work in the high entropy wind of core-collapse Supernovae 
\citep[see, e.g., ][]{farouqi2009}. In recent years, magneto-rotational supernovae have 
been identified as potential candidates \citep[see, e.g., ][]{nishimura2017}. 
Last year, the $r$-process went one step ahead of the game thanks to the recent neutron 
star merger GW170817. Characteristics of its electromagnetic counterpart
AT~2017gfo are  consistent with the hypothesis that during the merger event $r$-process elements were 
synthesized and ejected into ISM \citep[see, e.g.][and references therein]{2017Natur.551...80K}. 

During the \spr, instead, multiple neutron captures on a single seed 
nucleus (mostly \isotope{56}{Fe}) rarely occur. This is due to the fact that the 
$\beta$-decay half-life of freshly synthesized isotopes is shorter than the typical 
neutron capture timescale. Thus, the \spr\ mainly develops along the 
$\beta$-stability valley. Its products are commonly subdivided into a {\it main} 
component and a {\it weak} component. The first is synthesized in the interiors of 
Asymptotic Giant Branch (AGB) stars 
\citep{gallino1998,2006NuPhA.777..311S,cristallo2015}, 
while the latter is mainly produced during the core-He burning and shell-C burning 
phases of massive stars \citep{prantzos1990,chieffi2013}. 
The main neutron source in AGB stars is the $^{13}$C($\alpha$,n)$^{16}$O reaction, 
while the \nean reaction dominates in massive stars. 
Due to its rather low neutron densities ($n_n\approx 10^6$ cm$^{-3}$), the 
{\it weak} \spr\ cannot overcome the barrier opposed by 
nuclei with neutron magic number N=50: \isotope{88}{Sr}, \isotope{89}{Y}, and
\isotope{90}{Zr}. Thus, its major products are elements between the iron peak 
and the first \spr\ peak. 

In recent years, an {\it intermediate} neutron capture process (the so-called $i$-process), 
firstly described by \citet{cowan1977}, attracted a large interest with a series of recent 
dedicated articles \citep[and references therein]{denissenkov2017,cote2018}.
This peculiar process mainly occurs when protons are mixed in a hot convective
environment, producing neutron densities as large as $n_n\approx (10^{14}\div10^{16})$ 
cm$^{-3}$. In the models published so far (proton ingestion in low-mass low-metallicity 
AGB stars; very late thermal pulses or born-again AGBs; WDs rapidly accreting hydrogen; etc.), 
neutrons are released by $^{13}$C+$\alpha$ reactions, considering the large amount of 
 \isotope{13}{C} synthesized via proton captures on the abundant \isotope{12}{C}. 

As already recalled in the introductory section, in the S102+030 system a large amount 
of \isotope{22}{Ne} is available and, as it is shown below, at least during the SF accretion 
regime, the temperature at the He-burning shell attains and exceeds $3\times 10^8$K, so that 
weak \spr\ nucleosynthesis can occur. Moreover, as during the 
He-flashes episodes the temperature becomes  $\apgt 5\times 10^8$K, the 
neutron densities become very large, typical for the \ip nucleosynthesis. 
It is worth recalling that during the SF regime the accretor undergoes recurrent RLOF episodes, 
so that at a certain moment, when the He-flash is strong enough, the ejected matter 
becomes enriched in 
$n$-rich isotopes.

In order to properly describe the nucleosynthesis and the energy generation in the 
flashing S102+030 system we adopted in our computation a large nuclear network, 
including isotopes from hydrogen to the heaviest stable ones (\isotope{209}{Pb} and \isotope{210}{Bi}).
The original network was introduced by \citet{2006NuPhA.777..311S} to follow standard \spr;  
successively, it was extended by \citet{cristallo2011} to account for very large neutron exposure. 
In the present work the nuclear network includes about 700 isotopes linked by 1000 nuclear processes, 
including $\alpha-,\ p-,\ n-$captures as well as $\beta$-decays 
\citep[see also][]{cristallo2016}.

\section{The onset of nuclear burning}
\label{sec:beg}

Initial parameters of the accreting CO WD chosen in the present study are $\rm M_{CO,i}=$1.02047 \msun, 
central temperature and density $\rm T_C=1.71\times 10^7$~K and $\rm \rho_C=4.082\times 10^7$~\gcc,
respectively, $\rm \log(L/L_\odot)= -2.03$ and $\rm \log(T_{eff})= 4.31$. At the onset
of the mass transfer process the orbital separation of components is 
$a$=0.0614\rsun,
corresponding to the orbital period  $\simeq2.21$~min. 
We assume, as in Paper~II, that the expansion of the accreting WD is limited by its Roche lobe and 
excess of the matter is expelled from the system with the specific angular momentum of the accretor.
   \begin{figure}
   \centering
   \includegraphics[width=\columnwidth]{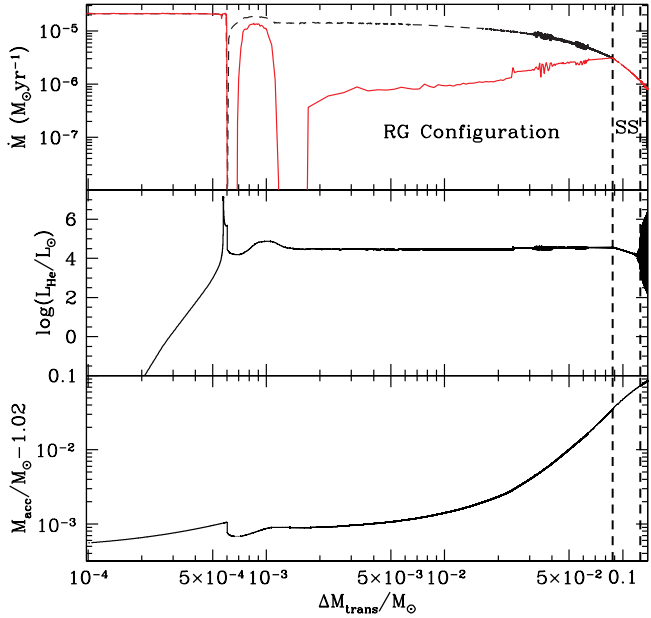}
   \caption{Evolution of the S102+030 system during the first two He-flash
   episodes and during the ``RG Configuration'' regime. Upper panel: mass
   transfer rate from the donor (dashed line) and effective mass deposition rate onto
   the accretor (solid line).}
              \label{figure1}
    \end{figure}

Evolution of the S102+030 system at the beginning of the mass transfer process is quite similar 
to that of the S102+020 (see Paper II), though, there exist some important differences. 
Figure~\ref{figure1} shows the evolution of the S102+030 system during this early phase
of mass transfer.  When, due to GW emission, He WD fills its Roche lobe, mass transfer begins at a 
very high rate ($\mdotd \sim 2.12\times 10^{-5}$\,\myr), close to the Eddington limit, while in model 
S102+020 it was an order of magnitude lower (\mdotd$\sim 2.2\times 10^{-6}$\,\myr). 

As a consequence, in the model S102+030 thermal energy at the surface of the accreting WD is delivered at a definitively larger rate, 
so that the first flash occurs after $5.6\times 10^{-4}$ \msun\ of He-rich matter 
has been accreted, while in the model S102+020 He-rich buffer piled up before the first 
flash was substantially more massive ($\sim 2.03\times 10^{-3}$\msun). 
At the onset of the He-flash, matter at the base of the He-rich buffer is less degenerate 
in model S102+030 than in model S102+020, so that the flash itself is less strong.
This is evident when considering the maximum luminosity during the He-flash $\rm L_{He}^{max}$, which is 
equal to $\rm 1.43\times 10^7 L_\odot$ and $\rm 3.74\times 10^9 L_\odot$ in models S102+030 and S102+020, respectively.
   \begin{figure}
   \centering
   \includegraphics[width=\columnwidth]{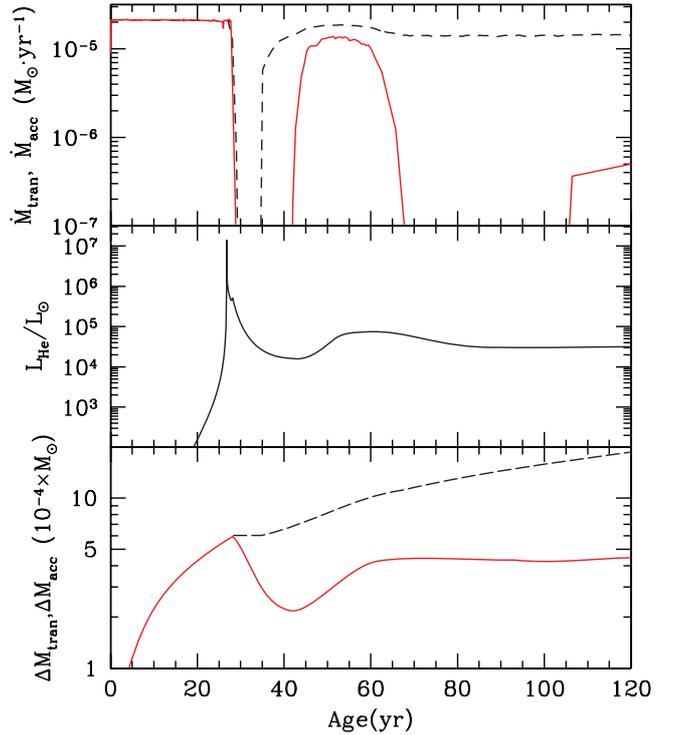}
   \caption{Evolution of the S102+030 system during the first two He-flash episodes. 
               The same quantities as in Fig.~\ref{figure1} are reported, but as a 
               functions of the time elapsed from the onset of mass transfer. In the lower 
               panel the amounts of matter transferred from the donor $\rm \Delta M_{tran}$ 
               (dashed line) and the matter deposited onto the accreting CO WD 
               $\rm\Delta M_{acc}$ (solid line) are plotted.}
              \label{figure2}
    \end{figure}

Energy release by the first thermonuclear flash results in the expansion of the accreting WD 
which fills its Roche lobe. Formation of the contact system leads to mass and angular 
momentum loss from the system. As a consequence the orbital separation increases and the 
donor detaches from its Roche lobe (Figs.~\ref{figure1}, \ref{figure2}). 
In binaries of lower mass (see Paper~II), the He WD donor fills again its Roche lobe after 
the He-flash died-down as, due to the GW emission, the binary system shrinks. 
In the more massive system under consideration, mass transfer from the donor resumes 
very soon, when He-burning is still providing the entire surface luminosity (see Table~\ref{tab1}). 
Such an occurrence is related to the fact that the amount of matter lost during 
the RLOF episode is quite small so that the system does not 
expand too much. As a 
consequence, when $\simeq$ 51\% of the previously accreted matter has been lost 
from the system (i.e. $\rm \Delta M_{lost}\simeq 3\times 10^{-4}$\msun), GW emission 
determines the onset of a new mass transfer episode. At that epoch, the accretor is still filling its 
Roche lobe, so that all the matter transferred from the donor is lost from the system and also 
the most external layers of the CO WD are eroded (see upper and lower panels in Fig.~\ref{figure2}). 
When He-burning efficiency definitively decreases, the accretor can start  to receive mass again. However, 
as \mdot\, from the donor is still very high (\mdotd$\simeq 2.09\times 10^{-5}$\msun\pyr), 
the further evolution of the S102+030 system occurs in the RG configuration accretion regime. 
After accreting additional $10^{-4}$\,\msun\ the CO WD experiences a second very mild He-flash 
($\rm L_{He}^{max}\sim 3 L_{star}$), which represents an additional engine to the expansion of the He-rich layers.
As a matter of fact, in the following evolution the whole mass transferred from the donor as well 
as part of the external layers of the accretor are lost by the system. 
At variance with the first He-flash episode, the system does not detach at all so that mass is 
transferred from the donor continuously. When the He-burning luminosity decreases to 
$\rm \log(L_{He}/L_\odot)$=4.466, the accretor starts to increase its mass again. 

As matter is transferred from the donor, the mass transfer rate decreases, but remains 
for some time in the range ($10^{-5}-10^{-6}$) \myr. 
However, accreting WD can burn steady only less than $ several \times  10^{-6}$\myr; 
excess of the matter should form an extended red-giant-like (RG) envelope (Paper~I). 
This means that the compact system is losing mass, being immersed into this envelope. 
We assume that spiral-in does not happen, since excess of the matter is minuscule, only 
several hundredth of \msun, and the matter is lost from the system. 
Really, during the RG configuration regime the transferred mass
is equal to $8.684\times 10^{-2}$\msun, while the mass effectively accreted is
equal to $3.415\times10^{-2}$\msun; hence, the average retention efficiency is
39.3\%. 

Next, the system enters steady-burning regime, when the matter transferred to the accretor 
is partially burned at $\mdot \sim 10^{-6}$ \myr\ and partially accumulated in 
the non-expanding envelope.

We neglected above the effects of tidal heating and possible tidal Novae 
 occurring when systems are close to contact 
\citep[e.g., ][]{1998ApJ...503..344I,2012ApJ...756L..17F}, since hydrogen-rich surface layers of massive post-common-envelope CO WD, which may be involved, are expected to have very low mass,
$\simeq 3\times 10^{-5}$\,\msun\ \citep{2018MNRAS.480.1547L}.
\section{The physical evolution of S102+030 system}
\label{sec:ph_evol}

The evolution during the SS, MF and SF regimes does not deserve any particular
analysis as compared to model S102+020 in Paper II: the mass transfer rate from 
the donor decreases, the mass ratio (mass of the primary over mass of the secondary)
increases and the separation of components increases. 
   \begin{figure}
   \centering
   \includegraphics[width=\columnwidth]{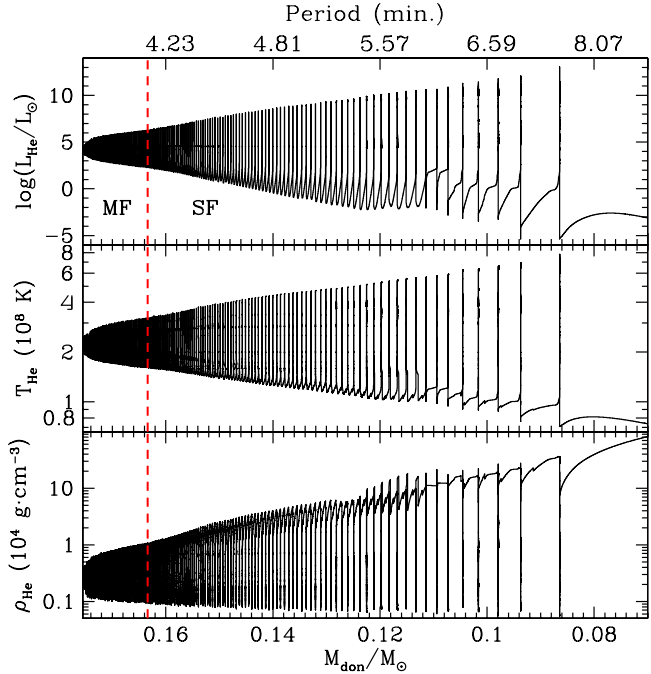}
   \caption{Physical properties of He-burning shell of the accretor of the
system S102+030 during the MF and SF regimes. From bottom to top: density, 
temperature the energy 
delivered per unit time via He-burning. Red heavy
dashed lines mark the transition from one burning regime to the other.
Lower x-axis: mass of the donor in solar masses; Upper x-axis: orbital period in minutes.
}
              \label{figure3}
    \end{figure}

In Fig.~\ref{figure3} we plotted the main physical characteristics of the
He-burning shell. 
Accreting CO WD enters the MF regime when its mass 
is 1.0927\msun\, and the donor mass is 0.1751\msun. At this epoch, the mass 
transfer rate is equal to \mdot=$1.152\times 10^{-6}$ \msunyrm. By comparing 
$\rm M_{acc}$ and \mdot\ values with those derived in Paper I, we find that the transition 
from the SS to the MF regime occurs at a slightly lower \mdot\, for a fixed $\rm M_{acc}$, 
due to the gravothermal energy  delivered by the compression of the C-rich layers piled-up 
via He-burning. The onset of the SF regime occurs when $\rm M_{acc}\simeq$1.1047\msun, 
and \mdotd=$8.14\times 10^{-7}$\myr. Accreting WD experiences 87 He-flashes, 
progressively stronger as $\rm M_{acc}$ increases and \mdotd\, decreases. 
When mass transfer resumes after the last He-flash, the mass transfer rate from the donor is 
\mdotd=$4.10\times 10^{-8}$\myr and the mass of the accretor is $\rm M_{acc}$=1.1312\msun. 
By extrapolating results in Paper I, if the mass transfer rate would be constant, a dynamical 
He-flash should occur after $\rm \Delta M_{He}\sim 0.04$\msun\, has been accreted. But, 
as already observed in Paper II, in binary systems with He-WD donor \mdotd\, 
continuously decreases and, hence, $\rm \Delta M_{He}$ becomes larger and larger as 
$M_{don}$ reduces. As a matter of fact we find that all the residual mass is transferred 
to the CO WD without triggering a new He-flash. Hence the final outcome of 
the evolution of S102+030 system is  formation of a massive WD, having a CO core of 
$\sim$1.1173\msun\, capped by a He-rich layer of 0.1004\msun.

Small ``irregularities'' in the curves for $\rm T_{He}$ and $\rm \rho_{He}$ between outbursts, 
that become visible in Fig.~\ref{figure3} for $\rm M_{don} \aplt 0.13$\,\msun, reflect change 
in the position of $\epsilon_{3\alpha,max}$, which depends on temperature, 
density and He abundance.

As it is well known, and also displayed in Fig.~\ref{figure3}, He-flashes become stronger 
as the mass transfer rate decreases. This implies that, pulse by pulse, the RLOF episode 
following each He-flash occurs sooner, while
the He-burning shell is still delivering a 
huge amount of energy, driving a very rapid expansion of the accreting WD. As 
a consequence the mass loss rate from the accretor becomes progressively larger, 
In particular we find that the RLOF episode triggered by the last He-flash lasts for 
21.92 yr and that $\simeq 6.9\times 10^{-3}$\msun\ is lost,
the mass loss rate exceeding 1\myr at the beginning of the flash.
   \begin{figure}
   \centering
   \includegraphics[width=\columnwidth]{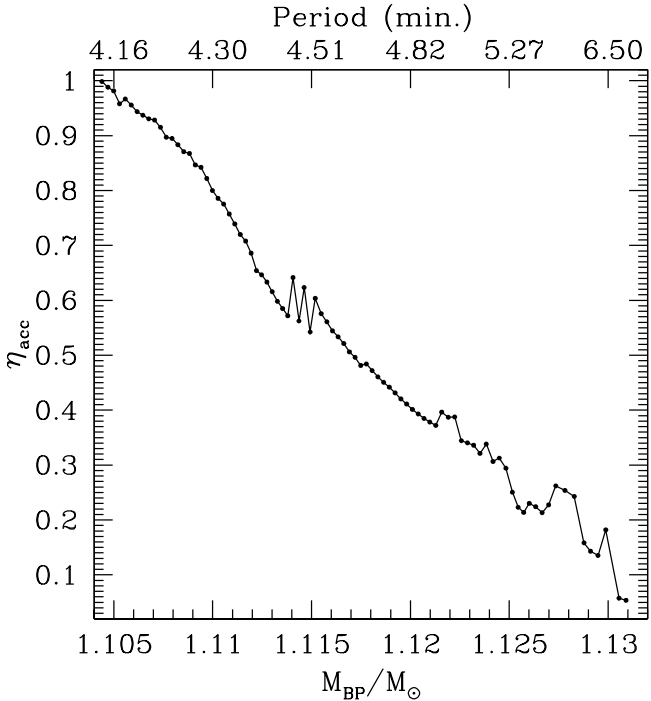}
   \caption{Retention efficiency during the SF regime as a function of the
            accretor mass at the bluest point along the loop in the HR
            diagram (lower x-axis) and of the orbital period (upper x-axis). }
              \label{figure4}
    \end{figure}

For each He-flash occurring in the SF accretion regime we derive the 
retention efficiency along the cycle $\eta$, defined as in Paper~I:
\begin{equation}
\mathrm{
\eta=1-{\frac{\Delta M_{L}}{\Delta M^1_{tr}+\Delta M^2_{tr}}},
}
\label{e:acef}
\end{equation}
where $\mathrm{\Delta M_{L}}$ is the mass lost during the RLOF, while
$\mathrm{\Delta M^1_{tr}}$ and $\mathrm{\Delta M^2_{tr}}$ are the mass accreted
onto the WD before and after the RLOF episode, respectively.
The obtained values 
are reported in Fig.~\ref{figure4} as a function of the accretor total mass at the bluest 
point along each loop in the HR diagram.
   \begin{figure}
   \centering
   \includegraphics[width=\columnwidth]{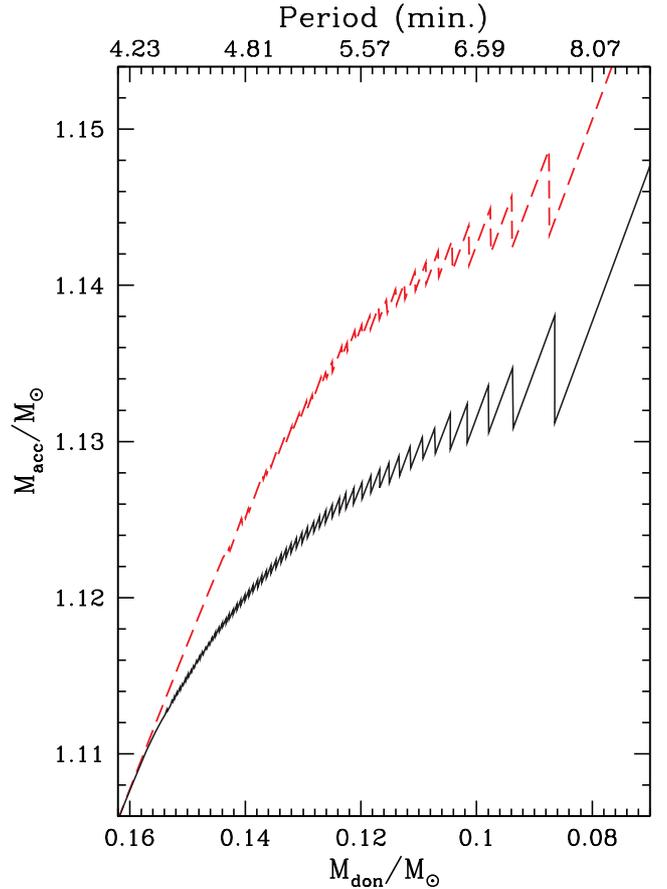}
   \caption{Solid line --- mass of the accretor as a function of the donor mass during 
     the SF accretion regime. Dotted red line --- the same quantity, 
     but for a model computed neglecting the nucleosynthetic and energetic contribution
     coming from $n$-captures. Upper x-axis reports the orbital period (in minutes).}
              \label{figure5}
    \end{figure}

As discussed in Paper II, the  value of $\eta$ reduces on average, as \mdotd\ 
decreases, even if the curve exhibits an irregular behavior. Once again, 
as we already suggested in the past (see Paper~II), such an occurrence has 
to be ascribed to the interplay of 1) decreasing mass transfer rate, 2) 
``dead-time'' span, between the end of RLOF and the re-onset of mass 
transfer, and 3) gravothermal energy delivered by the contraction of the 
freshly synthesized CO-rich layers below the actual He-mantle.
\begin{figure*}
\includegraphics[width=.48\textwidth,clip]{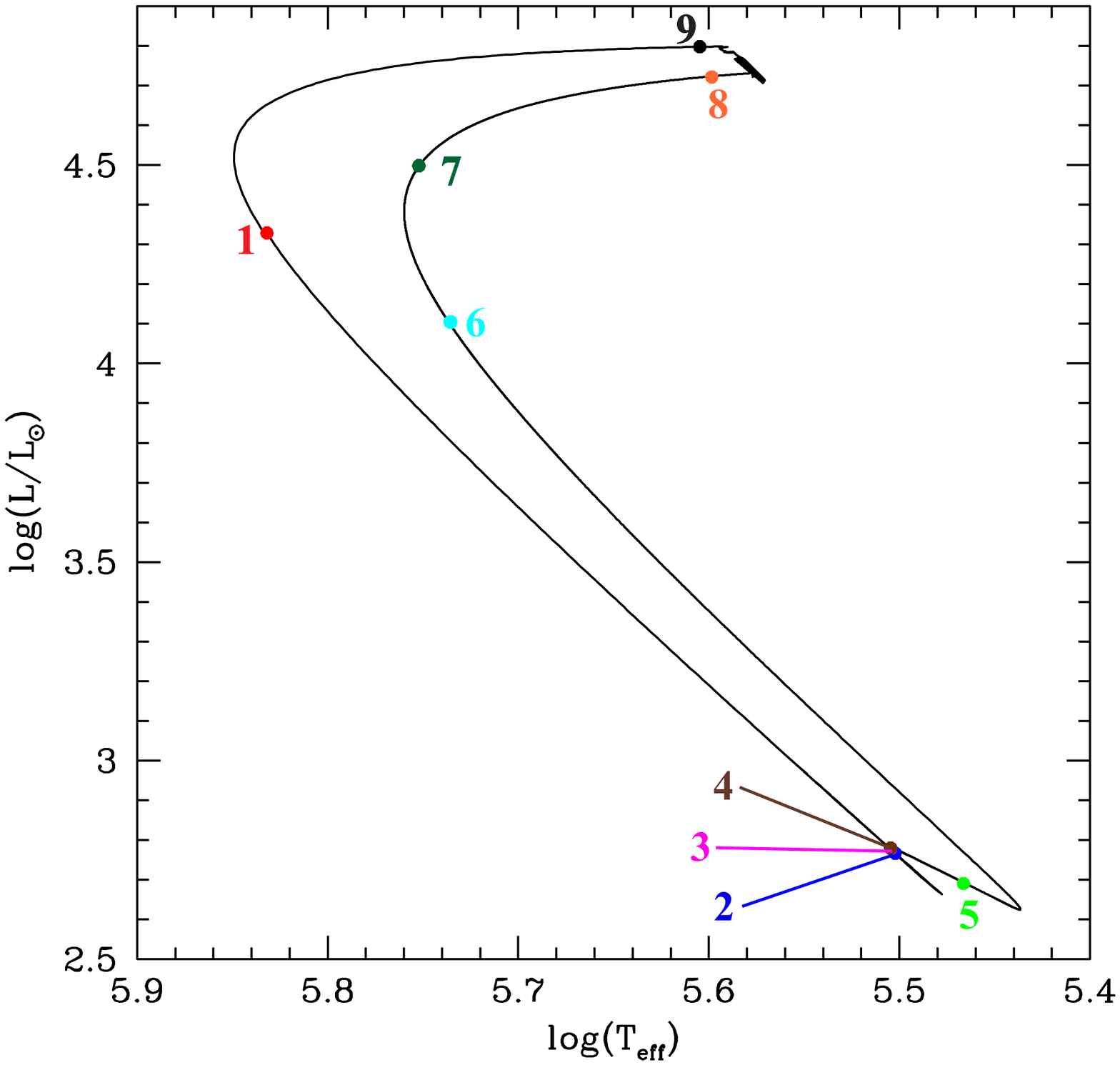}
\hskip 0.02\textwidth
    \includegraphics[width=.48\textwidth,clip]{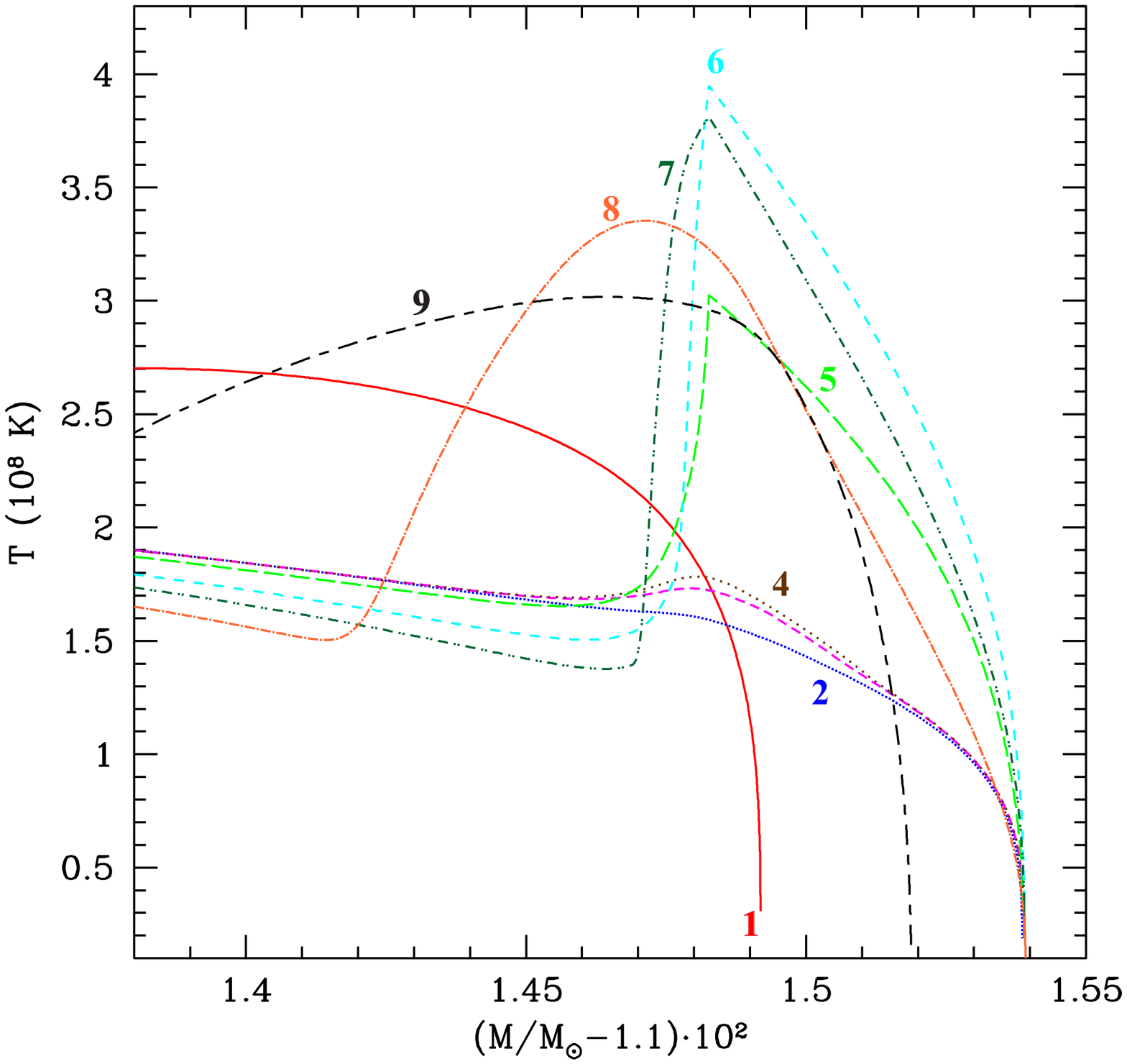}
   \caption{Left panel: evolution of  the accretor in the HR diagram during the 38$^{th}$ He-flash. 
Some relevant epochs are marked by filled dots and numbers.
Right panel: temperature profile as a function of the mass coordinate for the selected models in the 
 left panel, as labeled  inside the plot.}
\label{figure6}
\end{figure*}

A further inspection of Fig.~\ref{figure3} discloses that the temperature in the 
He-burning shell exceeds $3\times 10^{8}$ K already during the MF accretion 
regime, so that the \nean reaction becomes active. This demonstrates that in 
the computation of the evolution of S102+030 system a nuclear network 
involving all the relevant $n$-capture processes is needed to correctly evaluate 
both the nucleosynthesis and the delivered nuclear energy. 
To elucidate the importance of $n$-captures in the physical evolution of 
the S102+030 system we repeat the computation by neglecting all nuclear 
processes involving neutrons. In Fig.~\ref{figure5} we report the evolution 
of the accretor total mass  as a function of the donor total mass for 
both cases. As it can be noticed, when the contribution from $n$-captures
is accounted for (lower solid black curve), the accreting WD enters the SF 
regime earlier and the retention efficiency is lower
for the same mass lost by the donor.
Thus, when the contribution from $n$-captures is neglected (dotted red curve), 
a more massive WD is formed even if the final outcome of the accretion process
remains unchanged.  

Particular evolutionary stages considered above last for 
$\Delta T_{RG}=1.39\times 10^4$ yr,
$\Delta T_{SS}=2.39\times 10^4$ yr,
$\Delta T_{MF}=8.83\times 10^3$ yr,
$\Delta T_{SF}=5.65\times 10^5$ yr.
For the latter phase we refer to the time from the onset of the first strong He-flash (i.e., the epoch
corresponding to the bluest point along the loop in the HR diagram of the first strong He-flash)
to the resumption of mass transfer after the last (87$^{th}$) strong He-flash.

\begin{table*} 
\caption{Selected physical properties of the system S102+030 during the 38$^{th}$ He-flash episode in the Strong 
         Flashes accretion regime. The various epochs are the same displayed in Figs.~\ref{figure6} and \ref{figure7}. 
         The same quantities as in Table~\ref{tab1} are listed.}
\label{tab2} 
\centering 
  \begin{tabular}{l r r r r r r r r r}
   \hline\hline
{\it } & 1 & 2 & 3 & 4 & 5 & 6 & 7 & 8 & 9 \\
   \hline
$\mathrm{\Delta t}$ (yr)          &  ---     & 873.33   & 5.93     & 0.66     & 0.86     & 0.02     & 0.10     & 3.56     & 8.63     \\
$\mathrm{M_{acc}/M_\odot}$        & 1.114929 & 1.115397 & 1.115400 & 1.115400 & 1.115400 & 1.115400 & 1.115400 & 1.115402 & 1.115186 \\
$\mathrm{M_{don}/M_\odot}$        & 0.149851 & 0.149383 & 0.149380 & 0.149380 & 0.149379 & 0.149379 & 0.149379 & 0.149377 & 0.149377 \\
$a\mathrm{\ (in\ 10^{-2}R_\odot)}$& 9.7453   & 9.7649   & 9.7651   & 9.7651   & 9.7651   & 9.7651   & 9.7651   & 9.7652   & 9.7662   \\
$\mathrm{\log(T_{eff})}$          &   5.832  &   5.502  &   5.505  &   5.505  &   5.466  &   5.736  &  5.752   &   5.598  &  5.605   \\
$\mathrm{\log(L/L_\odot)}$        &   4.329  &   2.767  &   2.779  &   2.780  &   2.691  &   4.105  &  4.498   &   4.722  &  4.798   \\
$\mathrm{M_{He}/M_\odot}$         & 1.11478  & 1.1148   & 1.11481  & 1.11481  & 1.11483  & 1.11483  & 1.11483  & 1.11488  & 1.11498  \\
$\mathrm{\rho_{He}}$              & 3.932    & 24.277   & 16.676   & 16.637   & 9.395    & 6.235    & 1.571    & 0.851    & 1.686    \\
$\mathrm{T_{He}}$~(in $10^8$K)    & 2.057    & 1.608    & 1.643    & 1.687    & 3.025    & 3.948    & 3.812    & 3.153    & 2.820    \\
$\mathrm{\log(L_{He}/L_\odot)}$   & 3.218    & 3.091    & 3.749    & 3.975    & 7.619    & 9.034    & 6.451    & 4.939    & 4.540    \\
  \hline
  \end{tabular}
\end{table*}

\section{The chemical properties of the ejected matter}
\label{sec:che_evol}
We remind that, as already mentioned in the previous section, 
the temperature at the base of the He-burning shell exceeds 
$3\times 10^{8}$~K already during the MF phase. This implies that 
the \nean reaction is ignited very soon in the evolution of the system, 
so that during the He-flash neutrons are produced activating the 
corresponding nucleosynthesis of $n$-rich isotopes. However, we do 
not analyze this nucleosynthesis because during the MF regime no 
matter is lost from the accreting WD, so that there is no way to 
observe the imprint of these episodes. In fact, as demonstrated 
above, the S102+030 system does not produce any explosive 
event destroying the accreting WD.

\subsection{Heavy elements production in binary systems: a case study}

In order to illustrate how the nucleosynthesis proceeds during the SF 
regime, we focus our attention on the 38$^{th}$ He-flash episode, when 
the mass of the accretor at the re-onset of mass transfer from the donor is 
$\sim$1.1149\msun. We choose this episode because the physical properties during the He-flash and
the subsequent RLOF episode are such that the efficiency of $n$-capture nucleosynthesis is at 
a maximum. In particular the temperature in the He-rich buffer remains larger than the 
critical value for efficient neutrons production via \nean reaction for almost the entire time 
of the flash and of the subsequent RLOF episode. Moreover the convective shell 
triggered by the He-flash lasts long enough to guarantee a large neutron exposure (see below).

\begin{figure*}
   \centering
   \includegraphics[width=\textwidth]{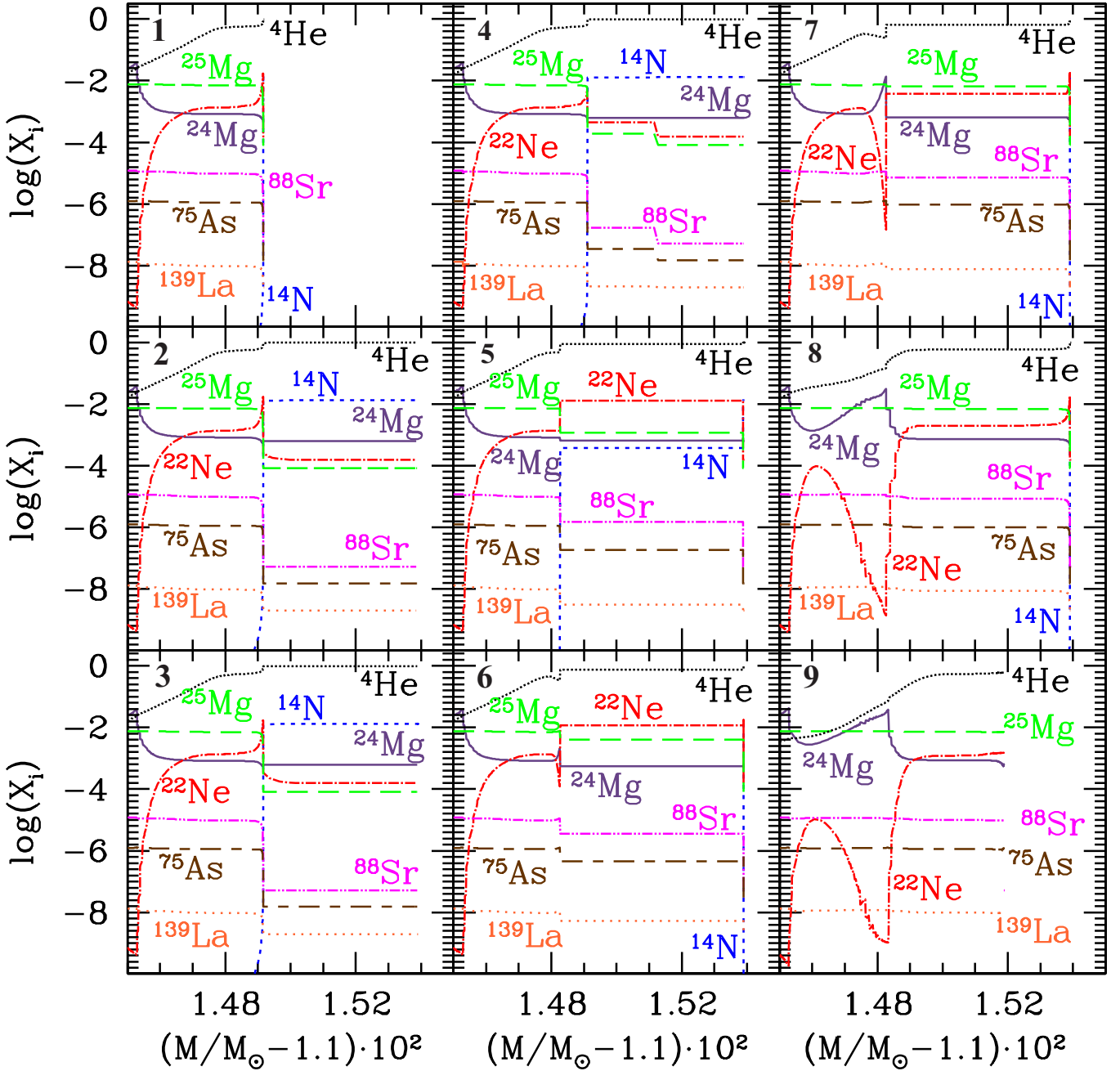}
   \caption{Abundances profile as a function of the mass coordinate of selected
isotopes for the accreting CO WD at the epochs marked in the left panel of 
Fig.~\ref{figure6}. Fine-dotted lines: \isotope{4}{He}; short-dashed lines:\isotope{14}{N}; 
dot-dashed lines: \isotope{22}{Ne}; solid lines: \isotope{24}{Mg}; long-dashed lines: \isotope{25}{Mg}; 
short-long dashed: \isotope{75}{As}; double-dotted dashed lines: \isotope{88}{Sr}; dotted lines: \isotope{139}{La}.}
   \label{figure7}
\end{figure*} 
In the left panel of  Fig.~\ref{figure6}\footnote{Colour version of Figs.6-12 are available in the online version.} 
we report the evolution of the accreting WD 
in the HR diagram; some relevant epochs are marked by heavy dots and numbers. The
temperature profiles in the outermost zone of the CO-core and in the He-rich
mantle for each of these models are plotted in the right panel of Fig.~\ref{figure6}, 
while the abundances of some key isotopes are reported in Fig.~\ref{figure7}. 
We select \isotope{4}{He} as a tracer of He-burning, 
\isotope{14}{N} and
\isotope{22}{Ne} as tracers of \nean reaction efficiency, 
\isotope{25}{Mg} as a tracer of \isotope{22}{Ne} destruction and neutrons production, 
\isotope{24}{Mg} as a tracer of the 
$\alpha$-capture reactions, 
\isotope{88}{Sr}, \isotope{75}{As}, and \isotope{139}{La} as tracers of $n$-capture nucleosynthesis. 
In Table~\ref{tab2} we list some physical properties of the accretor at the epochs 
marked
in Figs.~\ref{figure6} and \ref{figure7}.

 Point \textbf{1} in Fig.~\ref{figure6} corresponds to the resumption of mass
transfer from the donor. At that epoch the He-rich buffer is still hot, due to
the energy injected during the previous He-flash episode, even if thermal energy
is flowing inward, as clearly demonstrated by the broad shape of the
corresponding temperature profile in the right panel of Fig.~\ref{figure6}. 
As shown
in panel \textbf{1} of Fig.~\ref{figure7}, the surface abundances of
\isotope{4}{He} and \isotope{12}{C} are 0.835 and 0.133, 
respectively, clearly demonstrating the effects of the previous He-flash
episodes. 
Due to the inward thermal diffusion,  He-rich buffer piled-up via
mass deposition rapidly becomes  isothermal. 
After $\Delta t_{\rm 12} \approx 873$~yr, due to
the continuous mass deposition, the He-shell, located at $\sim 1.1148$\msun,
heats up to $1.608\times 10^8$~K, while the density increases up to $2.428\times
10^{4}$\gcc (see also the flat portion in the temperature profile labeled
\textbf{2} in the range $x=(M/M_\odot-1.1)\cdot 10^2=1.48-1.49$ in the right panel of 
Fig.~\ref{figure6}). At the epoch \textbf{2}, He-burning already delivers $L_{\rm He}\sim
1.23\times 10^3$ \lsun, a factor 2 larger than the surface luminosity of the
accreting WD. 
The increase of temperature in the He-rich buffer
determines also the burning of \isotope{14}{N} into \isotope{22}{Ne}, as 
it is seen by comparing the lowest parts of the lines for this isotope
in  panels \textbf{1} and \textbf{2} of Fig.~\ref{figure7} (at $x\sim 1.49$). 
After $\Delta t_{\rm 23}=5.93$ yr the temperature in the He-shell increases 
to $1.643\times 10^8$~K and He-burning
delivers $L_{\rm He}\simeq5.61\times 10^3$\lsun. At this epoch, He-burning is fully
ignited, even if no substantial nucleosynthesis has occurred so far\footnote{Let us 
remark that we define the He-burning ignition as the epoch when the energy
delivered via 3$\alpha$-reactions per unit time  exceeds by a factor 100 the surface
luminosity of the accreting WD.}. 
In the following evolution, from point
\textbf{3} to point \textbf {4} in the left panel of Fig.~\ref{figure6}, lasting
 for $\Delta t_{\rm 34}$=240 days, the thermonuclear runaway develops, thus triggering the
onset of a convective shell which grows in mass outward (see the jumps in the
\isotope{22}{Ne}, \isotope{25}{Mg}, \isotope{88}{Sr} and \isotope{75}{As}
profiles in panel \textbf{4} of Fig.~\ref{figure7}). Note that, as it is well
known, during the rapid increase of the energy locally delivered, the He-burning
shell and, as a consequence, the inner border of the convective shell driven by
the He-flash itself, move inward in mass (see Paper
II). This determines the enrichment of the He-rich zone with isotopes
synthesized during the previous He-flash episode (see the increase of
\isotope{22}{Ne}, \isotope{25}{Mg}, \isotope{88}{Sr} and \isotope{75}{As}
abundances in the range $x\sim1.49-1.51$). 
\begin{figure}
\centering
\includegraphics[width=\columnwidth]{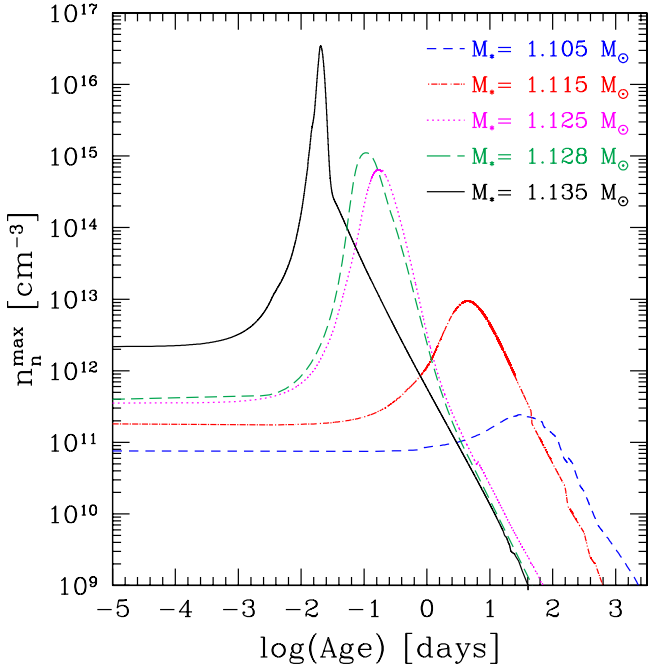}
\caption{Temporal evolution of the maximum neutron density attained during 
the 4$^{th}$ (short dashed line), 38$^{th}$ (dot-dashed line), 67$^{th}$ 
(dotted line), 76$^{th}$ (long-dashed line), and 87$^{th}$ (last) strong He-flashes (solid line).
For each flash, we fix the time origin at the epoch when 
the temperature at the He-burning shell first exceeds $3\times 10^8$ K.}
\label{figure8}
\end{figure} 

In the next $\Delta t_{\rm 45}=314$ days, the temperature in the
He-burning shell has reached $3.025\times 10^8$~K, and the 
convective shell has
practically attained its maximum extension (note the inward penetration
of the inner border of the convective shell by comparing panels \textbf{4} and
\textbf{5} in Fig.~\ref{figure7}). During this time span, \isotope{14}{N} has
been efficiently converted into \isotope{22}{Ne} in the entire convective shell.
In the following evolution, from \textbf{5} to \textbf{6}, lasting for 7.5 days,
$T_{\rm He-shell}$ continues to increase up to its maximum value over the pulse,
equal to $3.948\times 10^8$~K. \\
As the temperature at the He-shell exceeds the
threshold value for the activation of the \nean reaction, \isotope{22}{Ne} is
efficiently destroyed there (see the local minimum in the \isotope{22}{Ne}
profile in panel \textbf{6} of Fig.~\ref{figure7}), determining the onset of
neutron-isotopes nucleosynthesis. Isotope \isotope{25}{Mg} produced at the base of the
He-rich envelope as well as the products of $n$-capture processes (represented by
\isotope{88}{Sr}, \isotope{75}{As} and \isotope{139}{La} in Fig.~\ref{figure7})
are diluted over the whole convective region. When the accreting CO WD starts
to expand noticeably (at point \textbf{6} in the left panel of Fig.~\ref{figure6}),
thermal energy
starts to be transferred inward, so that, after $\Delta t_{\rm 67}\sim 34.9$~day,
the temperature in the He-burning shell decreases to $3.812\times 10^8$~K. 
The right panel in Fig.~\ref{figure6} also shows that
below the He-burning shell the temperature has increased above the threshold
value for the activation of \nean reaction so that the \isotope{22}{Ne}
abundance decreases also in the radiative zone between x=1.475 and x=1.48 (see
panel \textbf{7} in Fig.~\ref{figure7}). This process continues also during
the next $\Delta t_{\rm 78}\sim 3.565$ yr up to the onset of the Roche lobe overflow 
by the CO WD at \textbf{8}.
It is worth noticing that at the epoch \textbf{8}, \isotope{22}{Ne} has
been efficiently consumed up to $x\sim 1.46$. 
The RLOF episode lasts for $\Delta t_{\rm 89}\sim 8.627$ yr.
During RLOF, about $2.17\times 10^{-4}$\msun\ are lost,
corresponding to $\sim$ 45.7\% of the He-rich
matter transferred from the donor. 

As it is clearly shown by comparison of panel \textbf{2}, external part of panel \textbf{8} and panel \textbf{9}
in Fig.~\ref{figure7}, ejected matter is enriched, compared to ISM, by 
\isotope{4}{He} itself and  by the He-burning products, mainly \isotope{12}{C}, \isotope{16}{O},
\isotope{20}{Ne}, \isotope{24}{Mg}, by \isotope{25}{Mg} --- the tracer of \nean\ reaction, 
as well as by $n$-rich isotopes, traced by 
\isotope{88}{Sr},  \isotope{75}{As} and \isotope{139}{La.} Accreted \isotope{14}{N} is completely reprocessed.

\subsection{The overall nucleosynthesis path}

In order to understand and properly describe the nucleosynthesis processes occurring in the 
S102+030 system, it is useful to preliminary analyze the variation of neutron density during
recurrent strong He-flashes. Such a quantity $n_n$ is defined as 
\begin{equation}
n_n={\frac{X(n)}{A(n)}}\cdot \rho\cdot N_A,
\label{e:nn}
\end{equation}
where $\rho$ is the local density, $N_A$ is the Avogadro's number, $X(n), A(n)$ 
the abundance and the mass number of neutrons, respectively. 
A sizable $n$-capture nucleosynthesis can occur only if $n_n>10^9$ cm$^{-3}$.
In Fig.~\ref{figure8} we report the time evolution of the maximum neutron density 
$n_n^{max}$ for the 4$^{th}$, 38$^{th}$, 67$^{th}$, 76$^{th}$ and 87$^{th}$ 
He-flash episodes. Usually these quantities refer to the He-burning shell or to the layer 
above it, where \nean reactions occur.

For each selected model, we fix the time origin at the epoch when 
the temperature at the He-burning shell first exceeds $3\times 10^8$ K,
corresponding to 
the epoch when \nean reactions deliver neutrons at a sizable level.
\begin{table}
\centering 
\caption{From left to right we list the number of the strong He-flash, the value 
of the WD total mass (in \msun) at the bluest point along the HR loop, the maximum 
temperature (in $10^6$K) 
attained at the He-shell during the flash,  the time duration 
of the evolutionary phase during which $n_n^{max}>10^9$ cm$^{-3}$ and  
the total neutron exposure (in mbarn$^{-1}$).
 }
\label{tab3}
\begin{tabular}{rcccc}
\hline\hline
N. & M$_{BP}$ & T$_{max} [10^6 K]$ &  $\Delta$t & $\delta\tau$ [mbarn$^{-1}$]\\
\hline
 4  & 1.105 & 331 & 6.3 yr & 0.730 \\
38 & 1.115 & 395 & 1.7 yr & 2.317 \\
67 & 1.125 & 491 &  70.8 days & 3.162 \\
76 & 1.128 & 528 &  45.8 days & 4.300 \\
87 &1.135 &  762 & 40.2 days & 4.931 \\
\hline
\end{tabular}
\end{table} 

In Table~\ref{tab3} we report for the five selected models the mass of the accreting 
WD at the bluest point along the loop in the HR diagram, the maximum temperatures 
attained during the He-flash episode, the time span during which the maximum neutron 
density is larger than $10^9$ cm$^{-3}$ and 
the \textit{total neutron exposure}, 
 i.e. the time integrated neutron flux, defined as 
\begin{equation}
\delta\tau=\int_{\Delta t}n_n\cdot v_{th} dt,
\label{e:dtau}
\end{equation}
where $v_{th}=\sqrt{2k_B T/m_n}$ is the average thermal velocity of neutrons \citep{gallino1998}.

As it can be noticed, at the beginning of the SF accretion regime, when the WD mass 
is lower than $\sim 1.11$\msun, the maximum temperature attained during He-flashes 
is $\sim 3\times 10^8$ K, so that $n_n^{max}$ is lower than $3\times 10^{11}$ 
cm$^{-3}$ and nucleosynthesis in the \spr\ regime can occur. 
As the WD  mass increases, the temperature increases, attaining $\sim 7.6\times 10^8$ 
K in the last episode. However, the largest values for $n_n^{max}$  (above $\sim 10^{15}$ cm$^{-3}$), 
typical for the \ip\ regime are attained only in the last 11 He-flashes. 

At variance, the time over which the matter is exposed to high neutron densities 
(hereinafter \textit{exposure time}) 
decreases, as \mwd\, increases,  (and, thus, as the evolutionary time elapses).
Notwithstanding the total neutron exposure increases so that the production of heavy elements 
via $n$-captures should increase as the accreting WD continues to grow in mass.
\begin{table}
\centering 
\caption{Abundances of some selected isotopes in the matter accreted onto the CO WD (X$_{acc}$) and in 
the entire mass ejected during 87 strong He-flashes (X$_{ej}$).}
\label{tab4}
\begin{tabular}{rrrr}
\hline\hline
Isotope           &  Z &     X$_{acc}$        &     X$_{ej}$         \\    
\hline
\isotope{4}{He}   &  2 & 0.98                 & 4.98$\times 10^{-1}$ \\ 
\isotope{12}{C}   &  6 & 0.00                 & 3.80$\times 10^{-1}$ \\                              
\isotope{14}{N}   &  7 & 1.31$\times 10^{-2}$ & 4.27$\times 10^{-5}$ \\    
\isotope{16}{O}   &  8 & 0.00                 & 1.90$\times 10^{-2}$ \\    
\isotope{19}{F}   &  9 & 4.97$\times 10^{-7}$ & 3.16$\times 10^{-8}$ \\    
\isotope{20}{Ne}  & 10 & 1.94$\times 10^{-3}$ & 8.36$\times 10^{-3}$ \\    
\isotope{22}{Ne}  & 10 & 1.55$\times 10^{-4}$ & 5.09$\times 10^{-4}$ \\    
\isotope{23}{Na}  & 11 & 4.16$\times 10^{-5}$ & 5.51$\times 10^{-5}$ \\    
\isotope{24}{Mg}  & 12 & 6.24$\times 10^{-4}$ & 4.95$\times 10^{-2}$ \\    
\isotope{25}{Mg}  & 12 & 8.23$\times 10^{-5}$ & 1.32$\times 10^{-2}$ \\    
\isotope{26}{Mg}  & 12 & 9.42$\times 10^{-5}$ & 1.33$\times 10^{-2}$ \\    
\isotope{27}{Al}  & 13 & 7.22$\times 10^{-5}$ & 2.38$\times 10^{-4}$ \\    
\isotope{28}{Si}  & 14 & 8.11$\times 10^{-4}$ & 1.33$\times 10^{-2}$ \\    
\isotope{31}{P}   & 15 & 9.74$\times 10^{-6}$ & 1.03$\times 10^{-4}$ \\    
\isotope{32}{S}   & 16 & 4.17$\times 10^{-4}$ & 2.54$\times 10^{-4}$ \\    
\isotope{36}{S}   & 16 & 9.87$\times 10^{-8}$ & 4.74$\times 10^{-6}$ \\    
\isotope{40}{Ar}  & 18 & 0.00                 & 8.53$\times 10^{-6}$ \\    
\isotope{46}{Ca}  & 20 & 3.56$\times 10^{-9}$ & 2.81$\times 10^{-6}$ \\    
\isotope{45}{Sc}  & 21 & 4.90$\times 10^{-8}$ & 9.16$\times 10^{-7}$ \\    
\isotope{56}{Fe}  & 26 & 1.40$\times 10^{-3}$ & 1.66$\times 10^{-4}$ \\    
\isotope{60}{Fe}  & 26 & 0.00                 & 4.94$\times 10^{-4}$ \\    
\isotope{59}{Co}  & 27 & 4.15$\times 10^{-6}$ & 1.04$\times 10^{-4}$ \\    
\isotope{63}{Ni}  & 28 & 0.00                 & 8.04$\times 10^{-5}$ \\    
\isotope{64}{Ni}  & 28 & 8.96$\times 10^{-7}$ & 1.50$\times 10^{-4}$ \\    
\isotope{63}{Cu}  & 29 & 7.35$\times 10^{-7}$ & 3.25$\times 10^{-6}$ \\     
\isotope{65}{Cu}  & 29 & 3.38$\times 10^{-7}$ & 3.00$\times 10^{-5}$ \\     
\isotope{64}{Zn}  & 30 & 1.26$\times 10^{-6}$ & 1.11$\times 10^{-7}$ \\     
\isotope{69}{Ga}  & 31 & 4.84$\times 10^{-8}$ & 3.46$\times 10^{-6}$ \\     
\isotope{70}{Ge}  & 32 & 5.29$\times 10^{-8}$ & 3.67$\times 10^{-6}$ \\     
\isotope{75}{As}  & 33 & 1.52$\times 10^{-8}$ & 4.97$\times 10^{-7}$ \\     
\isotope{76}{Se}  & 34 & 1.54$\times 10^{-8}$ & 1.31$\times 10^{-6}$ \\     
\isotope{80}{Se}  & 34 & 9.06$\times 10^{-8}$ & 3.47$\times 10^{-6}$ \\     
\isotope{79}{Br}  & 35 & 1.50$\times 10^{-8}$ & 1.34$\times 10^{-7}$ \\     
\isotope{80}{Kr}  & 36 & 3.14$\times 10^{-9}$ & 2.05$\times 10^{-8}$ \\     
\isotope{86}{Kr}  & 36 & 2.65$\times 10^{-8}$ & 2.90$\times 10^{-6}$ \\     
\isotope{87}{Rb}  & 37 & 5.38$\times 10^{-9}$ & 1.59$\times 10^{-6}$ \\     
\isotope{86}{Sr}  & 38 & 6.14$\times 10^{-9}$ & 1.89$\times 10^{-7}$ \\     
\isotope{87}{Sr}  & 38 & 4.34$\times 10^{-9}$ & 3.22$\times 10^{-8}$ \\     
\isotope{88}{Sr}  & 38 & 5.23$\times 10^{-8}$ & 1.61$\times 10^{-6}$ \\     
\isotope{89}{Y}   & 39 & 1.31$\times 10^{-8}$ & 4.26$\times 10^{-7}$ \\     
\isotope{90}{Zr}  & 40 & 1.63$\times 10^{-8}$ & 4.70$\times 10^{-8}$ \\     
\isotope{96}{Zr}  & 40 & 9.47$\times 10^{-10}$& 2.21$\times 10^{-7}$ \\     
\isotope{138}{Ba} & 56 & 1.42$\times 10^{-8}$ & 2.08$\times 10^{-8}$ \\     
\isotope{139}{La} & 57 & 1.99$\times 10^{-9}$ & 2.54$\times 10^{-9}$ \\     
\isotope{208}{Pb} & 82 & 1.21$\times 10^{-8}$ & 2.78$\times 10^{-8}$ \\     
\hline                                           
\end{tabular}                                    
\end{table} 

\begin{figure}
   \centering
   \includegraphics[width=\columnwidth]{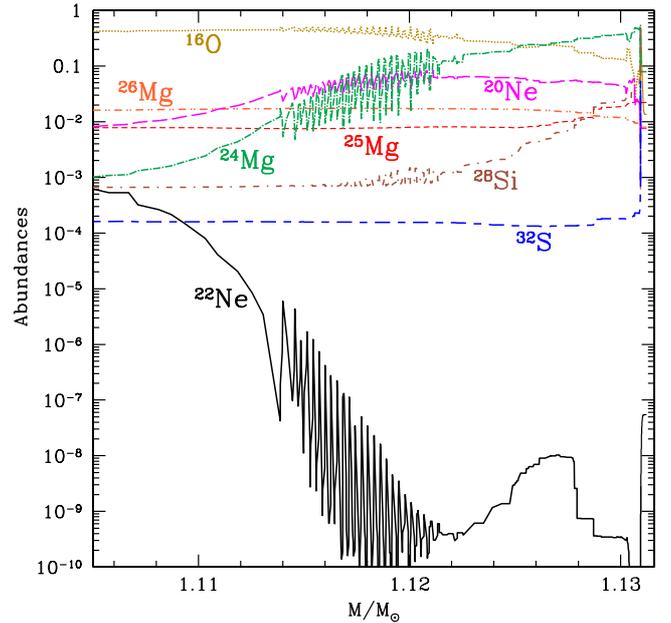}
   \caption{Abundances of selected light isotopes in the most external zone of the accreting WD, 
                piled up during the SF accretion regime. Dotted line: \isotope{16}{O}; long-dashed line:
\isotope{20}{Ne}; solid line: \isotope{22}{Ne}; dot-dashed line: \isotope{24}{Mg}; 
short-dashed line: \isotope{25}{S}; double-dot dashed line: \isotope{26}{Mg}; double-dash dotted line: 
\isotope{28}{Si}; long-short dashed line: \isotope{32}{S}.}
   \label{figure9}
\end{figure} 

\begin{figure}
\centering
\includegraphics[width=\columnwidth]{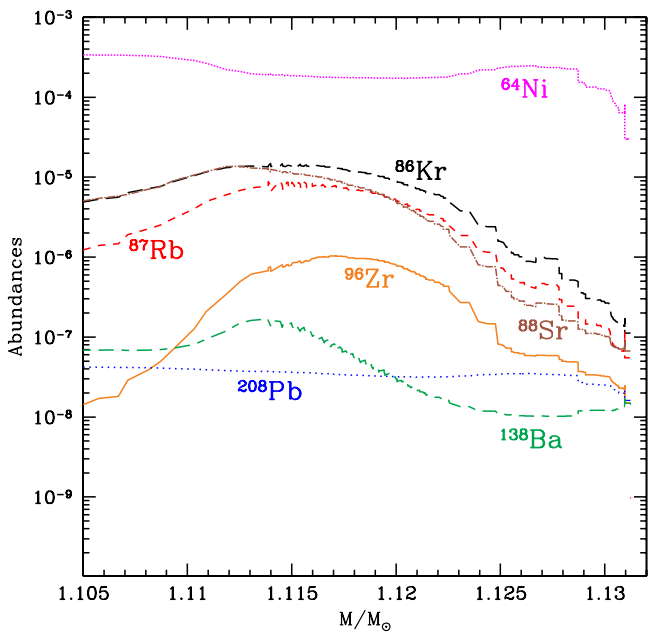}
\caption{Abundances of selected heavy isotopes, tracing the \spr\ 
nucleosynthesis occurring on the accreting WD during SF accretion regime.
Fine-dotted line: \isotope{64}{Ni}; long-dashed line: \isotope{86}{Kr}; 
short-dashed line: \isotope{87}{Rb}; dot-dashed line: \isotope{88}{Sr}
solid line: \isotope{96}{Zr}; long-short dashed line: \isotope{138}{Ba}; 
dotted line: \isotope{208}{Pb}.}
\label{figure10}
\end{figure}

\begin{figure}
\centering
\includegraphics[width=\columnwidth]{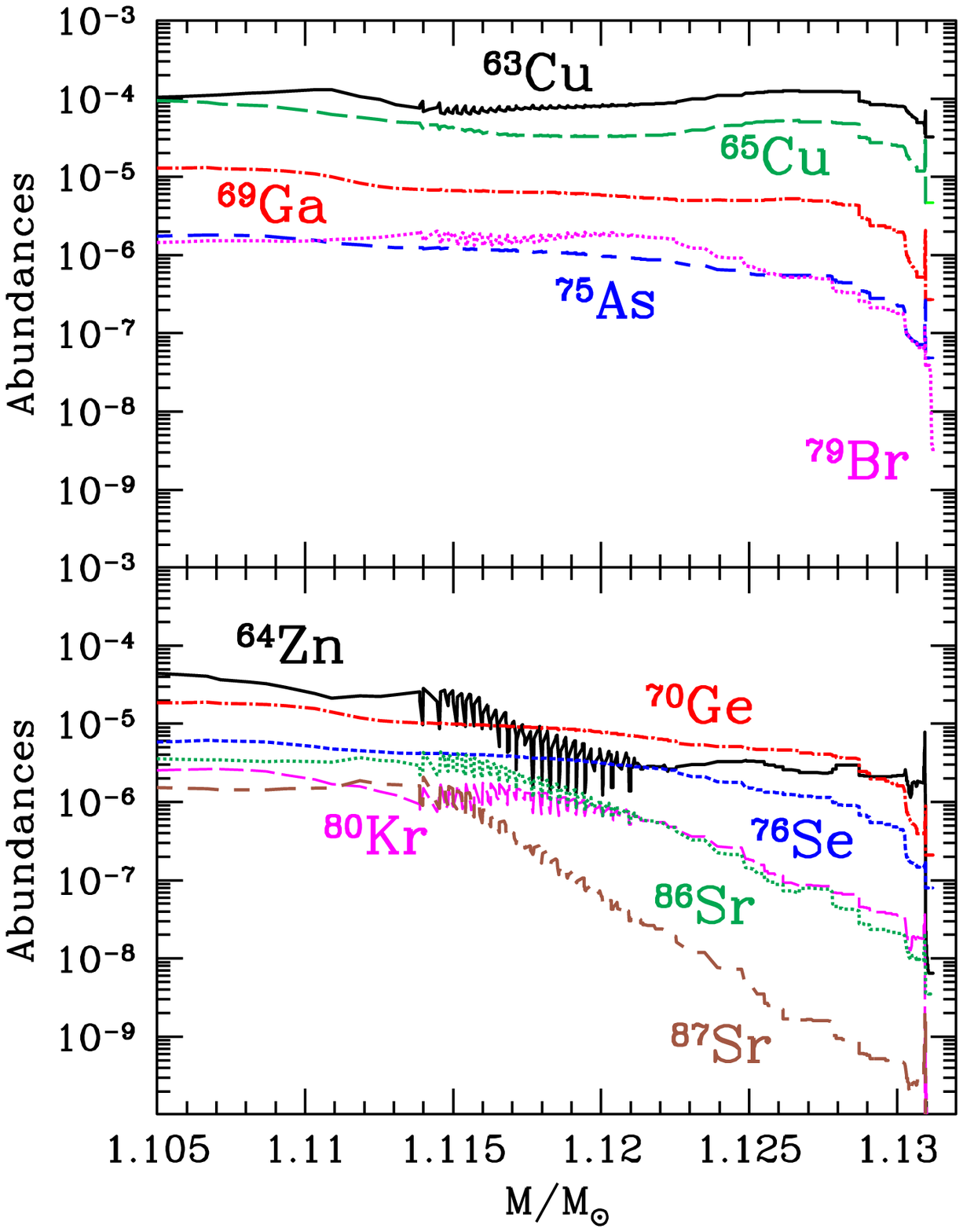}
\caption{As in Fig.~\ref{figure10}, but for isotopes normally ascribed to the {\it weak} 
           \spr\ (upper panel) and for $s$-only isotopes (lower panel).
        Upper panel: solid line: \isotope{63}{Cu}; long-dashed line: \isotope{65}{Cu}; 
dot-dashed line: \isotope{69}{Ga}; long-short dashed line: \isotope{75}{As}; 
dotted line: \isotope{79}{Br}. Lower panel: solid line: \isotope{64}{Zn}; dot-dashed line: \isotope{70}{Ge}; 
short-dashed line: \isotope{76}{Se}; long-dashed: \isotope{80}{Kr}; dotted line: \isotope{86}{Sr}; 
long-short dashed:\isotope{87}{Sr}.}
          \label{figure11}
\end{figure}

\begin{figure}
\centering
\includegraphics[width=\columnwidth]{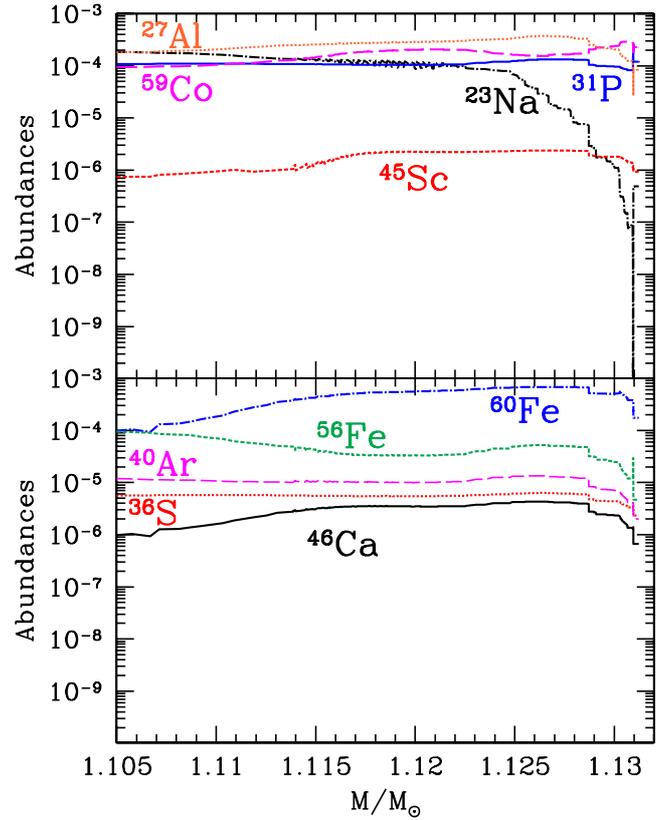}
\caption{As in Fig.~\ref{figure10}, but for light $n$-rich isotopes (lower panel) and 
light mono-isotopic elements (upper panel).
Upper panel: dot-dashed line: \isotope{23}{Na}; dotted line: \isotope{27}{Al}; 
solid line: \isotope{31}{P}; short-dashed line: \isotope{45}{Sc}; 
long dashed line: \isotope{59}{Co}. Lower panel: dotted line: \isotope{36}{S};  long dashed line: \isotope{40}{Ar}; 
short-dashed line: \isotope{56}{Fe}; dot-dashed line : \isotope{60}{Fe}.}
\label{figure12}
\end{figure}

In Figs.~\ref{figure9}-\ref{figure12} we plot the abundances 
of some selected isotopes in the outermost zones of the accreting WD after it has 
experienced RLOF triggered by the 87$^{th}$ (i.e. the last) He-flash, 
while in the third column  of Table~\ref{tab4} we list the abundance of the 
same isotopes in the accreted matter (X$_{acc}$).
The displayed layers have been piled up in the SF accretion regime so the chemical composition 
there traces the subsequent nucleosynthesis episodes. 
According to the discussion in the previous subsection, as soon as the 
temperature in the He-rich buffer increases above $3\times 10^{8}$ K during the flash episode, 
\isotope{22}{Ne} is consumed. Figure \ref{figure9} discloses that soon after entering 
the SF accretion regime, the temperature at the He-shell is so large that the \nean reaction 
is very efficient. This is evident when considering that \isotope{22}{Ne} abundance 
before the activation of this neutron source is determined mainly by the amount 
of \isotope{14}{N} in the accreted matter and, hence, it is as large as $2.08\times 10^{-2}$ 
(see Sect. \ref{sec:intro}). As a matter of fact, at the epoch of the 28$^{th}$ He-flash
episode, corresponding to the WD mass $M_{BP}\simeq 1.112$\msun, 
the \isotope{22}{Ne} abundance decreases to 
X(\isotope{22}{Ne})$\simeq 2\times 10^{-5}$. The saw-tooth behavior observed 
in Fig.~\ref{figure9} for 1.114 < M/\msun\ < 1.121 (corresponding to 
the 35$^{th}$ and 62$^{th}$ He-flash episodes) is determined by the fact that 
the evolutionary timescale during the flash episode becomes shorter and shorter so that 
the mixing efficiency in the He-rich buffer drastically decreases. As a consequence, during 
the flash, when the energy delivered by nuclear burning is still driving the increase 
of the He-shell temperature, the transport of \isotope{22}{Ne} down into the burning 
zones becomes less efficient. The local minima observed in Fig.~\ref{figure9} correspond 
to the position of the He-burning shell while the following local maxima are determined by 
the interplay between the low mixing efficiency 
during the He-flash and the quiescent 
He-burning after the RLOF episode. This sequence of events is well illustrated in Fig.~\ref{figure7}. 
The same effects define less expressed saw-tooth-like distribution of other isotopes.

Alpha-captures on \isotope{22}{Ne} produce \isotope{25}{Mg}, which 
acts as a strong neutron poison due to its large neutron capture cross-section \citep[see, e.g., ][]{massimi2017}; hence
\isotope{26}{Mg} is efficiently produced 
via the $^{25}$Mg(n,$\gamma$)$^{26}$Mg reaction.

Figure~\ref{figure9} also reveals that $\alpha$-isotopes are produced at a very high level. 
During the first He-flash episodes, when the maximum attained temperature is not very large, mainly,
\isotope{16}{O}, \isotope{20}{Ne} and \isotope{24}{Mg} are  
produced. 
After the 42$^{th}$ He-flash also \isotope{28}{Si} is efficiently produced.
Figure~\ref{figure9} also shows that \isotope{32}{S} is partially destroyed
(factor $\sim$ 2) pulse by pulse mainly via the 
\isotope{32}{S}(n,$\gamma$)\isotope{33}{S} reaction, while it is
produced after the 85$^{th}$  flash via $\alpha$-capture on 
\isotope{28}{Si}.
As a matter of fact, the maximum temperature attained during the SF accretion 
regime is not so high to allow the sizable production of heavier $\alpha$-isotopes. 
The production of \isotope{24}{Mg} plays a pivotal role in determining the $n$-capture 
nucleosynthesis because, like \isotope{25}{Mg}, it acts as a strong poison. 

In Fig.~\ref{figure10} we report the chemical profiles of some heavy isotopes,
representative of the nucleosynthesis of $n$-rich isotopes occurring in the He-rich buffer 
during the flash episodes. We plot \isotope{88}{Sr}, \isotope{138}{Ba} and
\isotope{208}{Pb} as tracers of the three peaks of the \spr. While
$^{88}$Sr is, on average, overproduced by a factor $\approx100$, the other two
isotopes show mild overabundance with respect to their initial values. This 
is somewhat expected because neutrons are released at the He-burning shell via 
\nean and are locally captured, while the new synthesized heavy isotopes are moved outward 
by convection, thus avoiding their local pile-up. As a consequence, abundance 
of elements beyond the first neutron magic number (N=50) is only marginally 
enhanced (by less than a factor 5).

In Fig.~\ref{figure10} we also report the profiles of some $n$-rich isotopes, whose
production in AGB stars is normally ascribed to the \nean reaction (e.g.
\isotope{86}{Kr}, \isotope{87}{Rb} and \isotope{96}{Zr}). For these isotopes, the
overproduction factors are even larger than for isotopes belonging to the three
peaks of the \spr\ (up to a factor \added{$\sim$} 1000 for \isotope{96}{Zr}). 
Figure~\ref{figure10} also discloses that, for M$\apgt$1.115\msun, abundances of all isotopes 
with A$>90$ start to decrease. Such an occurrence corresponds to the large production 
of neutron poisons, in particular \isotope{24}{Mg} (see Fig.~\ref{figure9}), 
which, as a matter of fact, largely reduces the  total neutron exposure 
available for $n$-capture nucleosynthesis. 
Isotopes with A$<90$ exhibit the same behavior, even if it occurs at larger 
masses (i.e. at later evolutionary time). In fact, the abundance of these isotopes 
is determined mainly by the peak neutron density and, hence, is less dependent on 
the neutron poison abundances. The reduced efficiency in producing these latter isotopes has 
to be ascribed to the fact that, as matter is accreted, the time duration of the
He-flash becomes shorter and, hence, also the duration of the peak phase of $n_n^{max}$  
becomes shorter (see Fig.~\ref{figure8}).
Neutron-rich isotopes with A$<$65 continue to be synthesized at a very high level, 
at least up to when a substantial neutron flux is available.  
For instance, \isotope{64}{Ni}, is overproduced 
by a factor $\sim$300 up to the 76$^{th}$ flash episode, corresponding to \mwd=1.1278\msun. 
Later on, neutron flux is reduced due to the efficient production of \isotope{28}{Si} and, less, 
of \isotope{32}{S} (see Fig.~\ref{figure9}). These $\alpha$-isotopes 
act as a strong neutron-poison, as their $n$-capture cross-section in the range (5-8)$\times 10^{8}$ K 
is comparable with that of \isotope{24}{Mg}. 

In the upper panel of Fig.~\ref{figure11} we report the profiles of isotopes normally 
ascribed to the {\it weak} \spr, demonstrating that the accretor in the S102+030 
system strongly overproduces  {\it weak}-$s$ elements. Among the others, 
\isotope{63}{Cu} exhibits a large overabundance 
due also to the important radiogenic contribution from \isotope{63}{Ni}.
In the lower panel of Fig.~\ref{figure11} we report the profiles of
some $s$-only isotopes whose production is shielded by any contribution from 
the $r$-process by their stable isobars (e.g. the couples \isotope{64}{Zn}-\isotope{64}{Ni}, 
\isotope{80}{Kr}-\isotope{80}{Se} and \isotope{87}{Sr}-\isotope{87}{Rb}). 
For all these isotopes we obtain very large overproduction factor 
at least up to when the increasing production of the \isotope{24}{Mg} poison 
as well as the reduction of the neutron exposure make their synthesis 
less efficient (see the decrease in the profiles starting at \mwd$\sim$1.115\msun).

As displayed in the lower panel of Fig.~\ref{figure12}, there are other 
neutron-rich isotopes whose production, contrarily to main \spr\ 
elements, continuously grows flash after flash. This is the case, for instance, of unstable isotope
\isotope{60}{Fe}, 
whose half-life is $\sim1.5\times 10^6$ yr. This isotope is of great interest 
for the astrophysical community, because it has been proved to be alive in 
the early solar system \citep{tang2012,mostefaoui2005}. The accreting WD 
in the S102+030 system  produces a large amount of \isotope{60}{Fe}, 
its abundance increasing up to $10^{-3}$. The same behavior is found for other 
$n$-rich isotopes, as \isotope{36}{S}, \isotope{40}{Ar} and \isotope{46}{Ca}. 
As a matter of fact, 
when neutron poisons are largely overabundant and/or neutron exposures are 
very low, only these isotopes show a significant production, in spite of the 
extremely large neutron densities attained.

At odds with the $n$-rich heavy elements nucleosynthesis 
discussed above,  the evolution of light elements abundance, as $^{19}$F 
and $^{23}$Na, holds few surprises (see upper panel in Fig.~\ref{figure12}). 
Fluorine is heavily destroyed, due to its high volatility in hot stellar interiors. 
Sodium  is destroyed at high temperatures via the \isotope{23}{Na}($\alpha$,p)\isotope{26}{Mg} 
reaction, whose cross section behaves similarly to that of the 
\isotope{24}{Mg}($\alpha$,$\gamma$)\isotope{28}{Si} (see the decrease 
of \isotope{23}{Na} abundance in the upper panel of Fig.~\ref{figure12}). Aluminum 
shows an almost zero production. 
On the contrary, phosphorus and scandium, which are mildly produced by 
standard \spr, exhibit a moderate overproduction by a factor $\sim$10 - 20.
Finally, cobalt shows a flat overproduction 
by a factor $\sim$15-20 all over the accreted layers. 
This is due to its vicinity to the main seeds for $n$-captures process,  i.e. 
\isotope{56}{Fe} 
nuclei,  which are almost uniformly 
destroyed over the whole time-span of SF accretion regime (see lower panel in 
Fig.~\ref{figure12}).

The chemical composition of the matter ejected pulse-by-pulse depends not only on the 
nucleosynthesis occurring during each He-flash, but also on the mass extension of the 
flash-driven convective shell and on the amount of mass ejected during the RLOF. 
In fact, heavy elements produced during the flash in the He-burning shell are dredged-up
 by convection. If the outermost zones unstable 
to convection are eroded by the RLOF-triggered mass loss, the ejected matter will be enriched in 
heavy elements. Of course, the stronger the He-flash, the more extended the convective shell, the larger the
mass lost during the RLOF and, hence,  larger the heavy elements abundance in the ejected matter. 
In particular the abundance of the $j$-isotope in the ejected matter can be expressed as:
\begin{equation}
X_j^{ej}={\frac{\Delta M_1}{\Delta M_{lost}}} X_j^{cv}+{\frac{\Delta M_2}{\Delta M_{lost}}} X_j^{acc}\,,
\label{e:xej}
\end{equation}
where $X_j^{cv}$ and $X_j^{acc}$ represent abundance of the $j$-isotope 
in the zone unstable for convection and that in the accreted matter, respectively.
 $\Delta M_{lost}=M_{WD}^1-M_{WD}^2$
is the amount of mass lost during the RLOF and $M_{WD}^1,\ M_{WD}^2$ are the WD mass at the onset and at the end 
of the RLOF episode, respectively. 

By defining $M_{max}^{cv}$ as the larger mass coordinate attained by convection during the 
flash episode,
it comes out that $\Delta M_1=\max{(0,M_{max}^{cv}-M_{WD}^2)}$ and 
$\Delta M_2=\min{(\Delta M_{lost},M_{WD}^1-M_{max}^{cv})}$. Equation~(\ref{e:xej}) clearly 
suggests that the heavy elements abundance in the ejected matter is expected to be lower 
than that in the surface layer of the WD after each flash. 
As pulse by pulse the convective unstable zone approaches the WD surface, the ratio $X_j^{ej}/X_j^{cv}$ increases 
up to its maximum value, i.e. 1.

To analyze the heavy elements abundance in the ejected matter we make use of the \textit{elemental 
enhancement factor} defined as
\begin{equation}
\Psi=\frac{N(el)_{ej}}{N(el)_{acc}}
\label{e:psi}
\end{equation}
where $N(el)_{ej}$ and $N(el)_{acc}$ are the number abundances of a given element in the 
ejected matter and in the accreted matter, respectively and are defined as
\begin{equation}
N(el)=\sum_i {\frac{X_i}{A_i}}\ ,
\label{e:nel}
\end{equation}
the summation being extended to all the isotopes with atomic number $Z_{el}$. 
By adopting the definition of $X_j^{ej}$ in Eq.~(\ref{e:xej}), the elemental enhancement factor can be written as
\begin{equation}
\Psi=\Psi_{cv}{\frac{\Delta M_1}{\Delta M_{lost}}} +{\frac{\Delta M_2}{\Delta M_{lost}}}, 
\end{equation}
where $\Psi_{cv}$ is the elemental enhancement factor in the zone unstable for convection. 
\begin{figure}
   \centering
   \includegraphics[width=\columnwidth]{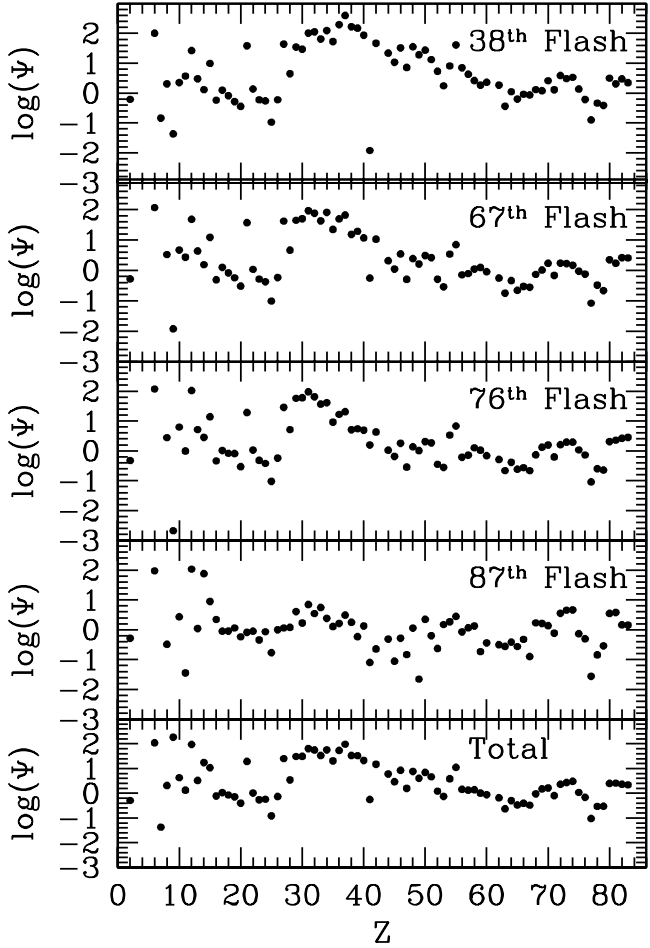}
   \caption{Elemental enhancement factor $\Psi$ in the matter ejected during the 38$^{th}$, 67$^{th}$, 76$^{th}$ and 87$^{th}$ 
                RLOF episodes, as labeled. The bottom panel shows the same quantity for the total mass ejected during the 
                SF accretion regime. }
  \label{figure13}
  \end{figure} 
As a matter of fact, we found that during the RLOF episodes triggered by the first  4 strong He-flashes the ejected matter is not 
enriched at all in heavy elements. 
In fact, the energy released by the flash is not very large so that the flash-driven convective zone remains confined in the innermost 
part of the He-rich layers and the mass lost during the RLOF episode is very low. 
Moreover the external border of the flash-driven convective shell attains the WD surface 
during the 42$^{th}$ flash, so that from this epoch up to the last He-flash $\Psi=\Psi_{cv}$. 
In Fig.~\ref{figure13} we plot (top to bottom) the $\Psi$ value for the ejected matter during the 38$^{th}$, 67$^{th}$, 
76$^{th}$ and
the 87$^{th}$ RLOF episodes; in the bottom panel we report the total elemental enhancement factor for the entire mass ejected during 
the SF accretion regime, equal to $M_{ej}^{tot}=5.012\times 10^{-2}$\msun. To evaluate such a quantity we assume that the matter 
ejected at each RLOF episode is fully mixed\footnote{This assumption is justified by the circumstance
that most matter is lost in the last $\sim10$ strong flashes, during which convective zone extends 
to the surface of accretor.}. 
Figure~\ref{figure13} shows that ejected matter is mainly enriched in $\alpha$-elements (up to \isotope{28}{Si}), 
by a factor 10-100 and in light-$s$ elements (Sr, Y, Zr). 
In the last column of Table~\ref{tab4} we list the abundances of selected isotopes in the matter ejected during the SF accretion regime (X$_{ej}$).

\citet{2013MNRAS.432.2048K} discovered an excess of Mg in the spectrum of a long-period
(52.96 $\pm$ 0.40 min.) AM CVn-type system. The origin
of excess should be associated with accretor,
since donor (respectively disc) matter does not have excess of Mg. \isotope{24}{Mg} is 
efficiently produced when the temperature in the He-burning layer increases above
$\sim 2.5\times 10^8$\,K. In the S102+030 system the outbursts in which \isotope{24}{Mg} may be produced
occur at $P_{orb}$ below 10 min. (see, e.g., Fig. 3 or 5), within about 1 Myr
after the formation of the system. Evolution to $P_{orb}\approx 53$ min. takes about 1.5 Gyr.
In this time, about 0.05\msun\, of matter non-enriched by magnesium is transferred to the accretor.
Outer layers of WD may become enriched in \isotope{24}{Mg} if, for instance, 
accreted matter mixes with underlying layers due to rotation.
We may expect that a similar process may occur in the AM CVn stars with He-star donors, since
accretors in them also experience outbursts with high enough temperature
\citep{2015ApJ...807...74B} and these systems also may evolve to long orbital periods
\citep{yungelson2008}.
But a more detailed study of this issue is beyond the scope of the present paper.

\section{Discussion and conclusion}
\label{sec:discus}
Present study is devoted to the investigation of physical and chemical evolution of
extreme ``interacting double-degenerate'' (IDD) AM CVn stars with both massive CO WD accretors
and He WD donors at the very early stage of their evolution (about first 60000 yr after formation
in the present case), when they have orbital periods below $\sim$ 10 min.
As in our previous study (Paper II) we model the donor as cool degenerate object obeying the \citet{zs69} mass-radius relation 
and adopt a time-dependent mass accretion rate as determined by angular momentum 
losses via emission of gravitational waves. We also assume that possible RLOF by the accreting WD 
triggers the loss of matter from the system with specific angular momentum of the accretor. 
In these systems the temperature attained in the He-burning shell during recurrent flash episodes 
is so large that a full nuclear network is necessary in order to correctly evaluate the energy contribution 
from $\alpha$-capture as well as from $n$-capture processes.

After the initial flash, that follows accumulation of the first layer of He susceptible to thermonuclear instability, 
as the mass transfer rate from the donor is continuously decreasing,
massive IDDs enter steady accretion regime. From this epoch on, the evolution of the 
S102+030 system occurs exactly as in systems with degenerate He WD of lower mass. 
In particular, we confirm our previous finding that accretors in IDDs end their evolution as CO WDs capped by 
a massive He-buffer, without developing the ``last'' very strong dynamical He-flash able to produce an 
SN .Ia \citep{bildsten2007}. It has to be noticed that, according to \citet{2015ApJ...807...74B}, also AM CVn 
stars of the ``He-star family'' never experience dynamical ``last'' He-flash.

At variance with systems considered in Paper II, we found that in IDD systems with 
massive donors the temperature in the He-shell exceeds $3\times 10^{8}$K already during the 
MF accretion phase. As a consequence, \nean reactions are fully ignited and $n$-capture nucleosynthesis 
can occur. We demonstrate that (see Fig.~\ref{figure5}) the inclusion of the energy coming from 
$n$-capture processes largely affects the retention efficiency during the SF accretion phase, 
determining a final CO core less massive as compared to the case when incomplete 
nuclear network is adopted in the computation. However, our results also show that the 
evolutionary outcome of massive AM CVn stars does not depend on the inclusion or not of $n$-capture processes.

The detailed computation of the evolution of the S102+030 system shows that during the SF accretion regime maximum 
temperature attained during He-flashes progressively increases from $\sim3\times 10^8$ K up to $\sim 7.6\times 10^8$ K, 
corresponding, respectively, to  maximum neutron densities of $10^{11}$ $\mathrm{cm^{-3}}$  (typical 
of \spr\ nucleosynthesis in intermediate 
mass AGB stars) and $\sim 3\times 10^{16}$ $\mathrm{cm^{-3}}$ (typical of \ip nucleosynthesis), respectively.
Even if the total neutron exposure largely increases pulse-by-pulse, the neutron flux available for 
the nucleosynthesis of $n$-rich heavy isotopes largely reduces after about 40 He-flash episodes. 
In fact, $\alpha$-isotopes produced by He-burning  (mainly \isotope{24}{Mg}, \isotope{28}{Si}) 
and their $n$-capture daughters nuclei act 
as strong neutron-poisons due to their very large abundances.
As a consequence, 
we found that during the first He-flash episodes elements belonging to all the three \spr\ peaks are overproduced by a 
large factor. Then, pulse by pulse, the enhancement factor $\Psi$ of elements belonging to the third and subsequently also to the second 
\spr\ peaks rapidly decreases to one. 

In summary, the nucleosynthesis in massive AM~CVn stars like the system S102+030 is expected to show 
the typical pattern of the {\it weak} component of \spr, superimposed to a tail extending to heavier elements 
(resulting from an efficient \isotope{22}{Ne} burning in a hot convective environment), with local peaks corresponding to very 
$n$-rich isotopes, tracers of the activation of a very high temperature $i$-process. 
To our knowledge, this is the first paper dealing with an \ip related to the $^{22}$Ne+$\alpha$ source.

It is interesting to note that the nucleosynthesis depends mainly on the metallicity of the He WD progenitor. In fact, 
the latter determines the amount of \isotope{14}{N} in the accreted matter and, hence, of neutrons delivered via \nean reactions.
As a consequence, AM CVn stars like S102+030 with low metallicity stellar progenitors are expected to
scarcely produce also elements belonging to the first $s$-peak, while those with large metallicity ($Z\ge2Z_\odot$) could 
produce efficiently also heavy-$s$ elements (Ba, La, Nd, Sm).
 
The S102+030 system ejected into the interstellar medium $\approx 5.2\times 10^{-2}$\msun\ 
with the same chemical composition of the accreted matter during the evolution in the RG-burning regime 
and $\approx 5.1\times10^{-2}$\msun\ of matter with enhanced abundance of $\alpha$- and $n$-rich isotopes.
(see Table~\ref{tab4} and Fig.~\ref{figure13}).

AM CVn stars similar to the S102+030 system are expected to be very rare, having  formation rate of 
$\sim 10^{-6}\div 10^{-5}$\pyr 
\citep{npv+01}. As a consequence, their contribution to the Galactic chemical evolution is, most probably, insignificant.

Though many instances of stellar evolution involving \ip have been investigated already,
for sure they are not exhaustive of the plethora of minor nucleosynthesis
events at work in stellar interiors. Whenever sufficient amounts of
\isotope{13}{C} or \isotope{22}{Ne} are stored in a stellar layer, in fact, a
new type of nucleosynthesis may develop. The efficiency of these processes
obviously depends on the temperature to which the material is exposed, on
environment conditions (radiative/convective) and on the timescale
of the evolutionary phase. Systems similar to S102+30 are an example of such ``peculiar''
nucleosynthesis sources.

\section*{Acknowledgments}
We acknowledge the referee for insightful comments.
We are indebted to G. Ramsey for providing us mass-transfer  rates based on \citet{2006ApJ...640..466B} mass-radius relation. LRY ackowledges hospitality and support of Rome and Teramo observatories and Max Planck Institute for Astrophysics (Garching) where a part of this study was accomplished.   



\bibliographystyle{mnras}
\bibliography{s102_030} 

\begin{thebibliography}{}
\makeatletter
\relax
\def\mn@urlcharsother{\let\do\@makeother \do\$\do\&\do\#\do\^\do\_\do\%\do\~}
\def\mn@doi{\begingroup\mn@urlcharsother \@ifnextchar [ {\mn@doi@}
  {\mn@doi@[]}}
\def\mn@doi@[#1]#2{\def\@tempa{#1}\ifx\@tempa\@empty \href
  {http://dx.doi.org/#2} {doi:#2}\else \href {http://dx.doi.org/#2} {#1}\fi
  \endgroup}
\def\mn@eprint#1#2{\mn@eprint@#1:#2::\@nil}
\def\mn@eprint@arXiv#1{\href {http://arxiv.org/abs/#1} {{\tt arXiv:#1}}}
\def\mn@eprint@dblp#1{\href {http://dblp.uni-trier.de/rec/bibtex/#1.xml}
  {dblp:#1}}
\def\mn@eprint@#1:#2:#3:#4\@nil{\def\@tempa {#1}\def\@tempb {#2}\def\@tempc
  {#3}\ifx \@tempc \@empty \let \@tempc \@tempb \let \@tempb \@tempa \fi \ifx
  \@tempb \@empty \def\@tempb {arXiv}\fi \@ifundefined
  {mn@eprint@\@tempb}{\@tempb:\@tempc}{\expandafter \expandafter \csname
  mn@eprint@\@tempb\endcsname \expandafter{\@tempc}}}

\bibitem[\protect\citeauthoryear{{Bildsten}, {Townsley}, {Deloye}  \&
  {Nelemans}}{{Bildsten} et~al.}{2006}]{2006ApJ...640..466B}
{Bildsten} L.,  {Townsley} D.~M.,  {Deloye} C.~J.,   {Nelemans} G.,  2006,
  \mn@doi [\apj] {10.1086/500080}, \href
  {http://esoads.eso.org/abs/2006ApJ...640..466B} {640, 466}

\bibitem[\protect\citeauthoryear{{Bildsten}, {Shen}, {Weinberg}  \&
  {Nelemans}}{{Bildsten} et~al.}{2007}]{bildsten2007}
{Bildsten} L.,  {Shen} K.~J.,  {Weinberg} N.~N.,   {Nelemans} G.,  2007,
  \mn@doi [\apjl] {10.1086/519489}, \href
  {http://adsabs.harvard.edu/abs/2007ApJ...662L..95B} {662, L95}

\bibitem[\protect\citeauthoryear{{Brooks}, {Bildsten}, {Marchant}  \&
  {Paxton}}{{Brooks} et~al.}{2015}]{2015ApJ...807...74B}
{Brooks} J.,  {Bildsten} L.,  {Marchant} P.,   {Paxton} B.,  2015, \mn@doi
  [\apj] {10.1088/0004-637X/807/1/74}, \href
  {http://esoads.eso.org/abs/2015ApJ...807...74B} {807, 74}

\bibitem[\protect\citeauthoryear{{Burbidge}, {Burbidge}, {Fowler}  \&
  {Hoyle}}{{Burbidge} et~al.}{1957}]{b2fh1957}
{Burbidge} E.~M.,  {Burbidge} G.~R.,  {Fowler} W.~A.,   {Hoyle} F.,  1957,
  \mn@doi [Reviews of Modern Physics] {10.1103/RevModPhys.29.547}, \href
  {http://adsabs.harvard.edu/abs/1957RvMP...29..547B} {29, 547}

\bibitem[\protect\citeauthoryear{{Chieffi} \& {Limongi}}{{Chieffi} \&
  {Limongi}}{2013}]{chieffi2013}
{Chieffi} A.,  {Limongi} M.,  2013, \mn@doi [\apj]
  {10.1088/0004-637X/764/1/21}, \href
  {http://adsabs.harvard.edu/abs/2013ApJ...764...21C} {764, 21}

\bibitem[\protect\citeauthoryear{{C{\^o}t{\'e}}, {Denissenkov}, {Herwig},
  {Ruiter}, {Ritter}, {Pignatari}  \& {Belczynski}}{{C{\^o}t{\'e}}
  et~al.}{2018}]{cote2018}
{C{\^o}t{\'e}} B.,  {Denissenkov} P.,  {Herwig} F.,  {Ruiter} A.~J.,  {Ritter}
  C.,  {Pignatari} M.,   {Belczynski} K.,  2018, \mn@doi [\apj]
  {10.3847/1538-4357/aaaae8}, \href
  {http://adsabs.harvard.edu/abs/2018ApJ...854..105C} {854, 105}

\bibitem[\protect\citeauthoryear{{Cowan} \& {Rose}}{{Cowan} \&
  {Rose}}{1977}]{cowan1977}
{Cowan} J.~J.,  {Rose} W.~K.,  1977, \mn@doi [\apj] {10.1086/155030}, \href
  {http://adsabs.harvard.edu/abs/1977ApJ...212..149C} {212, 149}

\bibitem[\protect\citeauthoryear{{Cristallo} et~al.,}{{Cristallo}
  et~al.}{2011}]{cristallo2011}
{Cristallo} S.,  et~al., 2011, \mn@doi [\apjs] {10.1088/0067-0049/197/2/17},
  \href {http://adsabs.harvard.edu/abs/2011ApJS..197...17C} {197, 17}

\bibitem[\protect\citeauthoryear{{Cristallo}, {Straniero}, {Piersanti}  \&
  {Gobrecht}}{{Cristallo} et~al.}{2015}]{cristallo2015}
{Cristallo} S.,  {Straniero} O.,  {Piersanti} L.,   {Gobrecht} D.,  2015,
  \mn@doi [\apjs] {10.1088/0067-0049/219/2/40}, \href
  {http://adsabs.harvard.edu/abs/2015ApJS..219...40C} {219, 40}

\bibitem[\protect\citeauthoryear{{Cristallo}, {Karinkuzhi}, {Goswami},
  {Piersanti}  \& {Gobrecht}}{{Cristallo} et~al.}{2016}]{cristallo2016}
{Cristallo} S.,  {Karinkuzhi} D.,  {Goswami} A.,  {Piersanti} L.,   {Gobrecht}
  D.,  2016, \mn@doi [\apj] {10.3847/1538-4357/833/2/181}, \href
  {http://adsabs.harvard.edu/abs/2016ApJ...833..181C} {833, 181}

\bibitem[\protect\citeauthoryear{{Denissenkov}, {Herwig}, {Battino}, {Ritter},
  {Pignatari}, {Jones}  \& {Paxton}}{{Denissenkov}
  et~al.}{2017}]{denissenkov2017}
{Denissenkov} P.~A.,  {Herwig} F.,  {Battino} U.,  {Ritter} C.,  {Pignatari}
  M.,  {Jones} S.,   {Paxton} B.,  2017, \mn@doi [\apjl]
  {10.3847/2041-8213/834/2/L10}, \href
  {http://adsabs.harvard.edu/abs/2017ApJ...834L..10D} {834, L10}

\bibitem[\protect\citeauthoryear{{Farouqi}, {Kratz}, {Mashonkina}, {Pfeiffer},
  {Cowan}, {Thielemann}  \& {Truran}}{{Farouqi} et~al.}{2009}]{farouqi2009}
{Farouqi} K.,  {Kratz} K.-L.,  {Mashonkina} L.~I.,  {Pfeiffer} B.,  {Cowan}
  J.~J.,  {Thielemann} F.-K.,   {Truran} J.~W.,  2009, \mn@doi [\apjl]
  {10.1088/0004-637X/694/1/L49}, \href
  {http://adsabs.harvard.edu/abs/2009ApJ...694L..49F} {694, L49}

\bibitem[\protect\citeauthoryear{{Fuller} \& {Lai}}{{Fuller} \&
  {Lai}}{2012}]{2012ApJ...756L..17F}
{Fuller} J.,  {Lai} D.,  2012, \mn@doi [\apjl] {10.1088/2041-8205/756/1/L17},
  \href {http://esoads.eso.org/abs/2012ApJ...756L..17F} {756, L17}

\bibitem[\protect\citeauthoryear{{Gallino}, {Arlandini}, {Busso}, {Lugaro},
  {Travaglio}, {Straniero}, {Chieffi}  \& {Limongi}}{{Gallino}
  et~al.}{1998}]{gallino1998}
{Gallino} R.,  {Arlandini} C.,  {Busso} M.,  {Lugaro} M.,  {Travaglio} C.,
  {Straniero} O.,  {Chieffi} A.,   {Limongi} M.,  1998, \mn@doi [\apj]
  {10.1086/305437}, \href {http://adsabs.harvard.edu/abs/1998ApJ...497..388G}
  {497, 388}

\bibitem[\protect\citeauthoryear{{Iben}, {Tutukov}  \& {Fedorova}}{{Iben}
  et~al.}{1998}]{1998ApJ...503..344I}
{Iben} Jr. I.,  {Tutukov} A.~V.,   {Fedorova} A.~V.,  1998, \mn@doi [\apj]
  {10.1086/305972}, \href {http://esoads.eso.org/abs/1998ApJ...503..344I} {503,
  344}

\bibitem[\protect\citeauthoryear{{Kasen}, {Metzger}, {Barnes}, {Quataert}  \&
  {Ramirez-Ruiz}}{{Kasen} et~al.}{2017}]{2017Natur.551...80K}
{Kasen} D.,  {Metzger} B.,  {Barnes} J.,  {Quataert} E.,   {Ramirez-Ruiz} E.,
  2017, \mn@doi [\nat] {10.1038/nature24453}, \href
  {http://adsabs.harvard.edu/abs/2017Natur.551...80K} {551, 80}

\bibitem[\protect\citeauthoryear{{Kupfer}, {Groot}, {Levitan}, {Steeghs},
  {Marsh}, {Rutten}  \& {Nelemans}}{{Kupfer}
  et~al.}{2013}]{2013MNRAS.432.2048K}
{Kupfer} T.,  {Groot} P.~J.,  {Levitan} D.,  {Steeghs} D.,  {Marsh} T.~R.,
  {Rutten} R.~G.~M.,   {Nelemans} G.,  2013, \mn@doi [\mnras]
  {10.1093/mnras/stt524}, \href
  {http://adsabs.harvard.edu/abs/2013MNRAS.432.2048K} {432, 2048}

\bibitem[\protect\citeauthoryear{{Lauffer}, {Romero}  \& {Kepler}}{{Lauffer}
  et~al.}{2018}]{2018MNRAS.480.1547L}
{Lauffer} G.~R.,  {Romero} A.~D.,   {Kepler} S.~O.,  2018, \mn@doi [\mnras]
  {10.1093/mnras/sty1925}, \href
  {http://esoads.eso.org/abs/2018MNRAS.480.1547L} {480, 1547}

\bibitem[\protect\citeauthoryear{{Marsh}, {Nelemans}  \& {Steeghs}}{{Marsh}
  et~al.}{2004}]{mns04}
{Marsh} T.~R.,  {Nelemans} G.,   {Steeghs} D.,  2004, \mn@doi [\mnras]
  {10.1111/j.1365-2966.2004.07564.x}, \href
  {http://ads.inasan.ru/cgi-bin/nph-bib_query?bibcode=2004MNRAS.350..113M&db_key=AST}
  {350, 113}

\bibitem[\protect\citeauthoryear{{Massimi} et~al.,}{{Massimi}
  et~al.}{2017}]{massimi2017}
{Massimi} C.,  et~al., 2017, \mn@doi [Physics Letters B]
  {10.1016/j.physletb.2017.02.025}, \href
  {http://adsabs.harvard.edu/abs/2017PhLB..768....1M} {768, 1}

\bibitem[\protect\citeauthoryear{{Mostefaoui}, {Lugmair}  \&
  {Hoppe}}{{Mostefaoui} et~al.}{2005}]{mostefaoui2005}
{Mostefaoui} S.,  {Lugmair} G.~W.,   {Hoppe} P.,  2005, \mn@doi [\apj]
  {10.1086/429555}, \href {http://adsabs.harvard.edu/abs/2005ApJ...625..271M}
  {625, 271}

\bibitem[\protect\citeauthoryear{{Nelemans}}{{Nelemans}}{2005}]{nelemans_amcvn05}
{Nelemans} G.,  2005, in {Hameury} J.-M.,  {Lasota} J.-P.,  eds, ASP Conf. Ser.
  330: The Astrophysics of Cataclysmic Variables and Related Objects. p.~27

\bibitem[\protect\citeauthoryear{{Nelemans}}{{Nelemans}}{2009}]{2009CQGra..26i4030N}
{Nelemans} G.,  2009, \mn@doi [Classical and Quantum Gravity]
  {10.1088/0264-9381/26/9/094030}, \href
  {http://adsabs.harvard.edu/abs/2009CQGra..26i4030N} {26, 094030}

\bibitem[\protect\citeauthoryear{Nelemans, Portegies~Zwart, Verbunt  \&
  Yungelson}{Nelemans et~al.}{2001}]{npv+01}
Nelemans G.,  Portegies~Zwart S.~F.,  Verbunt F.,   Yungelson L.~R.,  2001,
  {\aap}, 368, 939

\bibitem[\protect\citeauthoryear{{Nishimura}, {Sawai}, {Takiwaki}, {Yamada}  \&
  {Thielemann}}{{Nishimura} et~al.}{2017}]{nishimura2017}
{Nishimura} N.,  {Sawai} H.,  {Takiwaki} T.,  {Yamada} S.,   {Thielemann}
  F.-K.,  2017, \mn@doi [\apjl] {10.3847/2041-8213/aa5dee}, \href
  {http://adsabs.harvard.edu/abs/2017ApJ...836L..21N} {836, L21}

\bibitem[\protect\citeauthoryear{{Nomoto}}{{Nomoto}}{1982}]{nom82a}
{Nomoto} K.,  1982, \mn@doi [\apj] {10.1086/159682}, \href
  {http://ads.inasan.ru/cgi-bin/nph-bib_query?bibcode=1982ApJ...253..798N&db_key=AST}
  {253, 798}

\bibitem[\protect\citeauthoryear{{Panei}, {Althaus}, {Chen}  \& {Han}}{{Panei}
  et~al.}{2007}]{2007MNRAS.382..779P}
{Panei} J.~A.,  {Althaus} L.~G.,  {Chen} X.,   {Han} Z.,  2007, \mn@doi
  [\mnras] {10.1111/j.1365-2966.2007.12400.x}, \href
  {http://adsabs.harvard.edu/abs/2007MNRAS.382..779P} {382, 779}

\bibitem[\protect\citeauthoryear{{Piersanti}, {Tornamb{\'e}}  \&
  {Yungelson}}{{Piersanti} et~al.}{2014}]{2014MNRAS.445.3239P}
{Piersanti} L.,  {Tornamb{\'e}} A.,   {Yungelson} L.~R.,  2014, \mn@doi
  [\mnras] {10.1093/mnras/stu1885}, \href
  {http://adsabs.harvard.edu/abs/2014MNRAS.445.3239P} {445, 3239}

\bibitem[\protect\citeauthoryear{{Piersanti}, {Yungelson}  \&
  {Tornamb{\'e}}}{{Piersanti} et~al.}{2015}]{piersanti2015}
{Piersanti} L.,  {Yungelson} L.~R.,   {Tornamb{\'e}} A.,  2015, \mn@doi
  [\mnras] {10.1093/mnras/stv1452}, \href
  {http://adsabs.harvard.edu/abs/2015MNRAS.452.2897P} {452, 2897}

\bibitem[\protect\citeauthoryear{{Prantzos}, {Hashimoto}  \&
  {Nomoto}}{{Prantzos} et~al.}{1990}]{prantzos1990}
{Prantzos} N.,  {Hashimoto} M.,   {Nomoto} K.,  1990, \aap, \href
  {http://adsabs.harvard.edu/abs/1990A%26A...234..211P} {234, 211}

\bibitem[\protect\citeauthoryear{{Ramsay} et~al.,}{{Ramsay}
  et~al.}{2018}]{2018arXiv181006548R}
{Ramsay} G.,  et~al., 2018, preprint, \href
  {http://adsabs.harvard.edu/abs/2018arXiv181006548R} {} (\mn@eprint {arXiv}
  {1810.06548})

\bibitem[\protect\citeauthoryear{Savonije, de Kool  \& van~den Heuvel}{Savonije
  et~al.}{1986}]{skh86}
Savonije G.~J.,  de Kool M.,   van~den Heuvel E. P.~J.,  1986, \aap, 155, 51

\bibitem[\protect\citeauthoryear{{Shen}}{{Shen}}{2015}]{shen2015}
{Shen} K.~J.,  2015, \mn@doi [\apjl] {10.1088/2041-8205/805/1/L6}, \href
  {http://adsabs.harvard.edu/abs/2015ApJ...805L...6S} {805, L6}

\bibitem[\protect\citeauthoryear{{Solheim}}{{Solheim}}{2010}]{2010PASP..122.1133S}
{Solheim} J.,  2010, \mn@doi [\pasp] {10.1086/656680}, \href
  {http://adsabs.harvard.edu/abs/2010PASP..122.1133S} {122, 1133}

\bibitem[\protect\citeauthoryear{{Straniero}, {Gallino}  \&
  {Cristallo}}{{Straniero} et~al.}{2006}]{2006NuPhA.777..311S}
{Straniero} O.,  {Gallino} R.,   {Cristallo} S.,  2006, \mn@doi [Nuclear
  Physics A] {10.1016/j.nuclphysa.2005.01.011}, \href
  {http://adsabs.harvard.edu/abs/2006NuPhA.777..311S} {777, 311}

\bibitem[\protect\citeauthoryear{{Tang} \& {Dauphas}}{{Tang} \&
  {Dauphas}}{2012}]{tang2012}
{Tang} H.,  {Dauphas} N.,  2012, \mn@doi [Earth and Planetary Science Letters]
  {10.1016/j.epsl.2012.10.011}, \href
  {http://adsabs.harvard.edu/abs/2012E%26PSL.359..248T} {359, 248}

\bibitem[\protect\citeauthoryear{Tutukov \& Yungelson}{Tutukov \&
  Yungelson}{1996}]{ty96}
Tutukov A.~V.,  Yungelson L.~R.,  1996, {\mnras}, 280, 1035

\bibitem[\protect\citeauthoryear{Tutukov, Fedorova, Ergma  \&
  Yungelson}{Tutukov et~al.}{1985}]{tfey85}
Tutukov A.~V.,  Fedorova A.~V.,  Ergma E.~V.,   Yungelson L.~R.,  1985, \sval,
  11, 52

\bibitem[\protect\citeauthoryear{Verbunt \& Rappaport}{Verbunt \&
  Rappaport}{1988}]{vr88}
Verbunt F.,  Rappaport S.,  1988, {\apj}, 332, 193

\bibitem[\protect\citeauthoryear{{Wu}, {Wang}, {Liu}  \& {Han}}{{Wu}
  et~al.}{2017}]{2017A&A...604A..31W}
{Wu} C.,  {Wang} B.,  {Liu} D.,   {Han} Z.,  2017, \mn@doi [\aap]
  {10.1051/0004-6361/201630099}, \href
  {http://adsabs.harvard.edu/abs/2017A%26A...604A..31W} {604, A31}

\bibitem[\protect\citeauthoryear{{Yungelson}}{{Yungelson}}{2008}]{yungelson2008}
{Yungelson} L.~R.,  2008, \mn@doi [Astronomy Letters]
  {10.1134/S1063773708090053}, \href
  {http://adsabs.harvard.edu/abs/2008AstL...34..620Y} {34, 620}

\bibitem[\protect\citeauthoryear{Zapolsky \& Salpeter}{Zapolsky \&
  Salpeter}{1969}]{zs69}
Zapolsky H.~S.,  Salpeter E.~E.,  1969, {\apj}, 158, 809

\makeatother
\end{thebibliography}

\bsp	
\label{lastpage}
\end{document}